\documentclass[journal]{IEEEtran}
 \usepackage{booktabs}

\usepackage{amsmath}                
\usepackage{amsthm}                 
\usepackage{amssymb}
\usepackage{thmtools}
\usepackage{kpfonts}
\usepackage[geometry]{ifsym}
\usepackage{dsfont}
\usepackage{multirow}
\usepackage[mathscr]{euscript}
\usepackage{graphicx}
\usepackage{color}
\usepackage{paralist}
\usepackage{framed}
\usepackage{float}
\usepackage{longtable}
\usepackage{cite}
\usepackage[colorlinks=true, citecolor=magenta, linkcolor=blue]{hyperref}       
\usepackage[font=itshape]{quoting}

\usepackage{listings}
\usepackage{framed} 
\usepackage{nomencl} 
\makenomenclature

\declaretheoremstyle[%
  headfont=\bfseries,%
  headpunct={:},%
  notefont=\normalfont\bfseries,%
  notebraces={--~}{},
    qed=$\blacksquare$,
]{definitionstyle}
\theoremstyle{definition}
\declaretheorem[style=definitionstyle,name=Definition]{defn}

\declaretheorem[style=definitionstyle,name=Algorithm]{alg}

\theoremstyle{definition}

\theoremstyle{plain}
\theoremstyle{remark}

\begin{document}
%
\title{Hetero-functional Network Minimum Cost Flow Optimization: A Hydrogen-Natural Gas Network Example}
\author{Wester C.H. Schoonenberg\thanks{Thayer School of Engineering, Dartmouth College, Hanover, New Hampshire, USA, Wester.C.H.Schoonenberg.TH@Dartmouth.edu}, Amro M. Farid\thanks{Thayer School of Engineering, Dartmouth College, Hanover, New Hampshire, USA, amfarid@dartmouth.edu}}
\maketitle

\begin{abstract} 
Over the past decades, engineering systems have developed as networks of systems that deliver multiple services across multiple domains. This work aims to develop an optimization program for a dynamic, hetero-functional graph theory-based model of an engineering system. The manuscript first introduces a general approach to define a dynamic system model by integrating the device models in the hetero-functional graph theory structural model. To this end, the work leverages Petri net dynamics and the hetero-functional incidence tensor. The respective Petri net-based models are translated into the quadratic program canonical form to finalize the optimization program. The optimization program is demonstrated through the application of the program to a hydrogen-natural gas infrastructure test case. Four distinct scenarios are optimized to demonstrate potential synergies or cascading network effects of policy across infrastructures. 

This work develops the first hetero-functional graph theory-based optimization program and demonstrates that the program can be used to optimize flows across a multi-operand network, transform the operands in the network, store operands over time, analyze the behavior for a quadratic cost function, and implement it for a generic, continuous, large flexible engineering systems of arbitrary topology. 
\end{abstract}
\vspace{-0.2in}
\section{Introduction}\label{CH6:sec:Introduction}
Over the past decades, engineering systems have developed as networks of systems that deliver multiple services across multiple domains \cite{de-Weck:2011:01}. Examples of such socio-technical systems are the electrified transportation system\cite{Amini:2019:00,Bilgin:2015:00,Farid:2016:ETS-BC05}, the energy-water nexus\cite{Stillwell:2011:00,Hussey:2012:00,Bazilian:2011:00,Lubega:2014:EWN-J11}, and the multi-modal energy system\cite{Thompson:2020:00}. These systems have become increasingly interdependent across domains as a result of market forces and the associated pursuit of efficiency and cost reductions \cite{Schoonenberg:2018:ISC-BK04}. For example, the New England electric power grid relies more than ever on natural gas for its electricity generation, whereas the same natural gas is also needed to heat homes in the winter. 

The interdependence of engineering system services has lead to a need for a better understanding of the holistic dynamics and trade-offs in these systems \cite{de-Weck:2011:01,Crawley:2015:00}. Modeling tools can support the pursuit for more insight into engineering system and their optimal control. These tools need to be quantitative, represent the heterogeneity of the modeled system, and be generalizable across domains \cite{Schoonenberg:2020:ISC-BC11}. 

Existing optimization methods are generally based on conventional graph theoretic approaches, or on discipline and application specific dynamic models. Minimum cost flow programs, for example, are based on networks\cite{Newman:2009:00} and consequently fail to address heterogeneity of function. The multilayer networks community has aimed to expand graph theory to accommodate heterogeneity of function\cite{De-Domenico:2013:01}, but Kivela et. al. have identified eight modeling limitations to the types of systems that can be modeled with multi-layer networks \cite{Kivela:2014:00}. Consequently, optimization programs based on those foundations inherently impose those same limitations. A graph-based approach was also used in the multi-commodity network flow optimization programs\cite{Ishimatsu:2016:00,Ishimatsu:2017:00,Ishimatsu:2020:00}. This approach does implement a notion of heterogeneity of function, but it does not integrate a specific description of operand state or storage in its program. Finally, approaches that optimize discipline or application specific programs lack generalizability\cite{Schoonenberg:2020:ISC-BC11}.

Hetero-functional Graph Theory, however, provides a rigorous modeling method that does not impose the previously mentioned modeling limitations of multilayer networks\cite{Schoonenberg:2018:ISC-BK04}. Furthermore, hetero-functional graph theory has been used in a variety of engineering system applications, to define both structural \cite{Farid:2007:IEM-TP00,Farid:2015:IEM-J23,Farid:2015:ISC-J19,Farid:2017:IEM-J13,Khayal:2017:ISC-J35,Schoonenberg:2018:ISC-BK04,Thompson:2020:SPG-C68} and dynamic models\cite{Farid:2016:ETS-BC05,Farid:2016:ETS-J27,Schoonenberg:2015:IEM-C48,Schoonenberg:2017:IEM-J34,vanderWardt:2017:ETS-J33,Viswanath:2014:ETS-C33}. However, hetero-functional graph theory has not been used as a foundation to an optimization program. This work proposes the first hetero-functional graph theory-based optimization program.

\vspace{-0.2in}

\subsection{Original Contribution}
This work intends to define the first hetero-functional network minimum cost flow optimization program. This entails that the optimization program balances supply and demand of multiple types of operands at distinct locations over time. The work solves the problem as a linearly constrained, convex quadratic program. The program can be applied to a wide variety of unlike application domains, as the operands may be transformed, assembled, and disjoined.

In the process of developing the hetero-functional network minimum cost flow optimization program, this work also establishes the first formal connection between the hetero-functional incidence tensor, arc-constant colored Petri nets, and the engineering system net. Furthermore, it establishes the first integration of device models to the system service feasibility matrices that couple the engineering system net dynamics to the operand behavior. 

Finally, this work demonstrates the hetero-functional network minimum cost flow optimization program by optimizing the first hydrogen-natural gas infrastructure test case. 

\vspace{-0.2in}
\subsection{Outline}
The background (Sec. \ref{CH6:sec:Background}) provides an introduction to Hetero-functional Graph Theory and Petri nets. The former is used as the structural backbone of the model, and the latter is used as a foundation to describe the system's dynamics. Sec. \ref{CH6:sec:HFGTdynamics} introduces the hetero-functional graph based dynamic model that incorporates device models. Sec. \ref{CH6:sec:HFGTprogram} then defines the hetero-functional network minimum cost flow optimization program. Sec. \ref{CH6:sec:IllustrativeExample} introduces a hydrogen-natural gas networked infrastructure test case as an example engineering system. This test case is modeled and optimized in Sec. \ref{CH6:sec:Results}. Sec. \ref{CH6:sec:Results} presents the hetero-functional graph model, the minimum cost flow optimization program, and the outcomes of the optimization program for the specified test case. Finally, Sec. \ref{CH6:sec:Conclusion} concludes the work and recaps the main contributions of the work to the literature. 

\section{Background}\label{CH6:sec:Background}
Hetero-functional Graph Theory (HFGT) was introduced over a decade ago for the study of reconfigurability of manufacturing systems \cite{Farid:2007:IEM-TP00,Farid:2008:IEM-J04,Farid:2008:IEM-J05,Farid:2008:IEM-J06} and has since been applied to a number of large flexible engineering systems including electric power grids, water systems, transportation systems, healthcare, and interdependent infrastructures.  Schoonenberg et al. \cite{Schoonenberg:2018:ISC-BK04} have produced a consolidating text on Hetero-functional Graph Theory, which has been further extended to include a tensor-based formulation\cite{Farid:2020:ISC-JR06}.  Hetero-functional graph theory introduces a large number of modeling constructs that are not found in ``traditional" graph theory\cite{Schoonenberg:2018:ISC-BK04,Farid:2020:ISC-JR06}.  Therefore, in order to maintain the self-contained nature of this paper many of the prerequisite terms are defined here for the reader's convenience and will serve as the basis for developing the hetero-functional network dynamics in Sec. \ref{CH6:sec:HFGTdynamics} and the hetero-functional network minimum cost flow in Sec. \ref{CH6:sec:HFGTprogram}.  This section also introduces several relevant definitions from the Petri net literature \cite{Jensen:1992:00,Popova-Zeugmann:2013:00}.  More specifically timed arc-constant colored Petri nets serve as an intermediate modeling vehicle that facilitates the transformation of a hetero-functional graph into hetero-functional network minimum cost flow optimization program. 

This section starts with an overview of the System Concept in Hetero-functional Graph Theory in Sec. \ref{CH6:subsec:B:HFGTsysconcept}.  After which, it continues to discuss the hetero-functional incidence tensor in Sec. \ref{CH6:subsec:B:HFGTincidencetensor}. Sec. \ref{CH6:subsec:B:timedPN} then covers Timed Petri nets that are used in Sec. \ref{CH6:subsec:B:HFGTservices} as a foundation for the Hetero-functional Graph Theory Service Model. 
Sec. \ref{CH6:subsec:B:bag} introduces mathematical foundations for multi-sets (i.e. bags) which is required for the introduction of Arc-Constant Colored Petri nets in Sec. \ref{CH6:subsec:B:accpn}.

\vspace{-0.15in}
\subsection{Hetero-functional Graph Theory:  System Concept}\label{CH6:subsec:B:HFGTsysconcept}
The first hetero-functional graph theory modeling construct is the system concept.   
\begin{defn}[System Concept\cite{Farid:2006:IEM-C02,Farid:2007:IEM-TP00,Farid:2008:IEM-J05,Farid:2008:IEM-J04,Farid:2015:ISC-J19,Farid:2016:ISC-BC06}]\label{CH6:def:systemConcept}
A binary matrix $A_S$ of size $\sigma(P)\times\sigma(R)$ whose element $A_S(w,v)\in\{0,1\}$ is equal to one when action $e_{wv} \in {\cal E}_S$ (in the SysML sense) is available as a system process $p_w \in P$ being executed by a resource $r_v \in R$  The $\sigma()$ notation is used return the size of a set.
\end{defn} 

In other words, the system concept forms a bipartite graph between the set of system processes and the set of system resources\cite{Farid:2015:ISC-J19}. The definition of the system concept relies on several other definitions: system resource, system process, and system operand.  

\begin{figure}[h]
\centering
\includegraphics[width=2.8in]{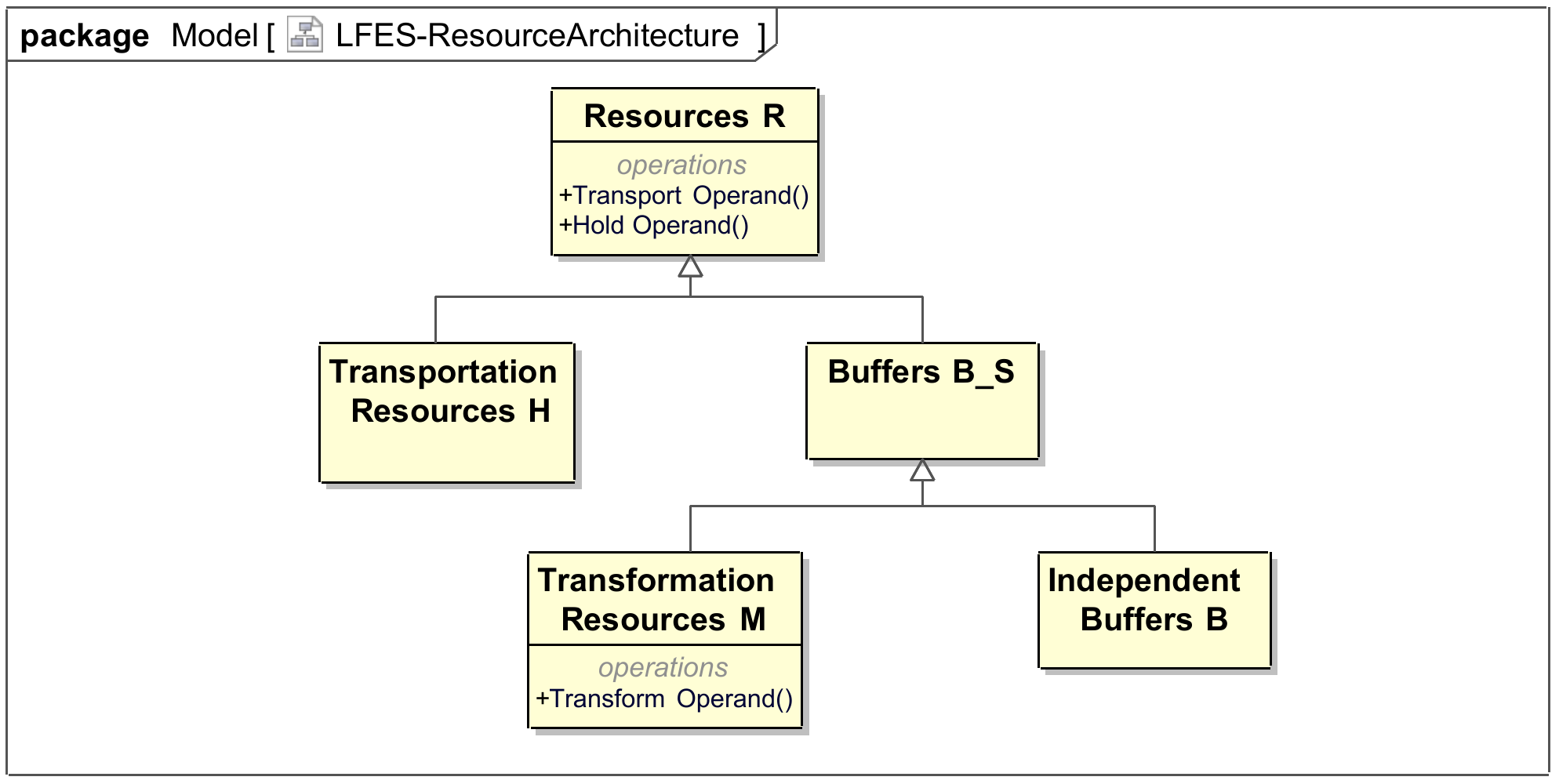}
\caption{A SysML Block Diagram: the meta-architecture of the allocated architecture of an LFES from a system form perspective\cite{Schoonenberg:2018:ISC-BK04}.}
\label{CH6:fig:HFGT:AA:BD}
\end{figure}

\begin{figure}[h]
\centering
\includegraphics[width=2.8in]{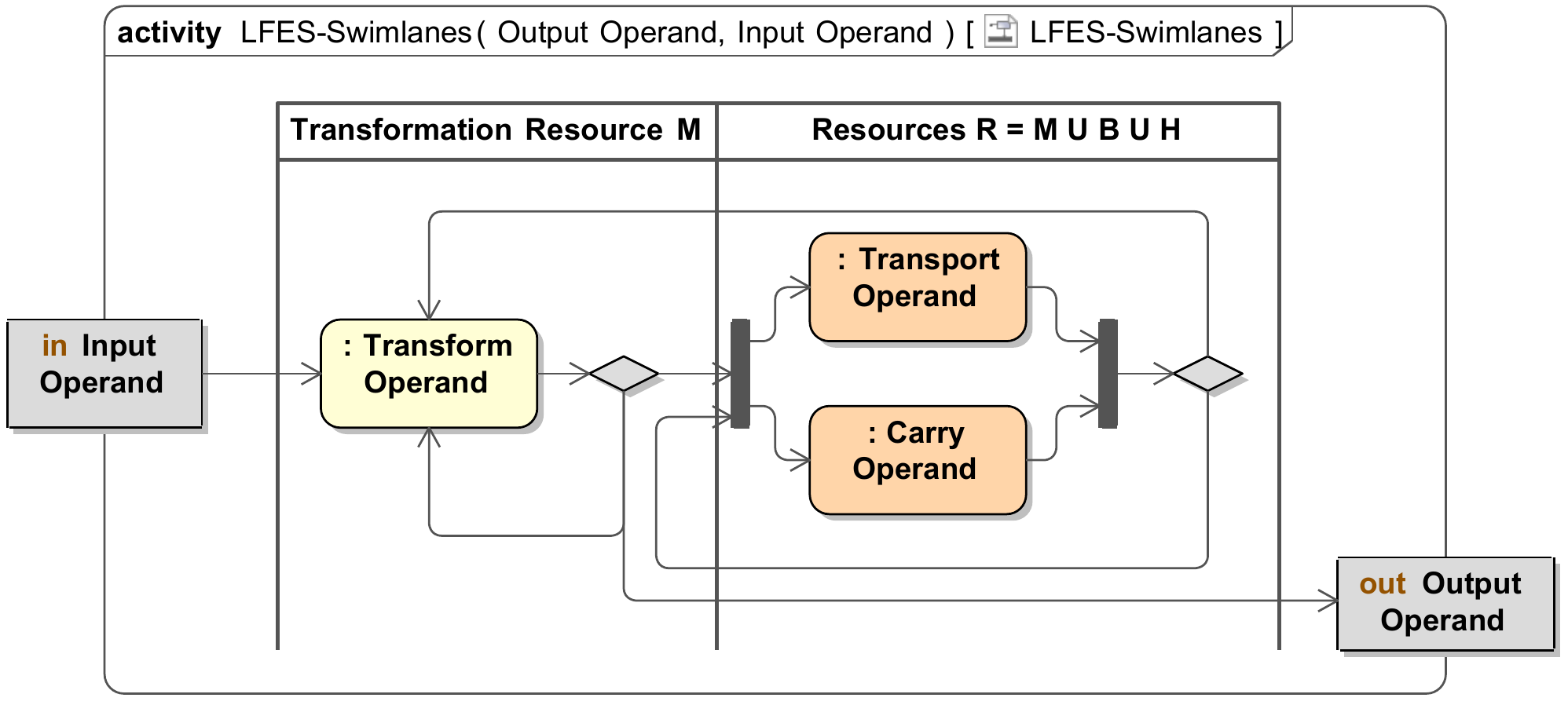}
\caption{A SysML Activity Diagram with swim lanes: the meta-architecture of the allocated architecture of an LFES from a system function perspective\cite{Schoonenberg:2018:ISC-BK04}.}
\label{CH6:fig:HFGT:AA:AD}
\end{figure}

\begin{defn}[System Resource]\cite{SE-Handbook-Working-Group:2015:00}
An asset or object $r_v \in R$ that is utilized during the execution of a process.  
\end{defn}
\begin{defn}[System Process\cite{Hoyle:1998:00,SE-Handbook-Working-Group:2015:00}]\label{CH6:def:CH4:process}
An activity $p \in P$ that transforms a predefined set of input operands into a predefined set of outputs. 
\end{defn}
\begin{defn}[System Operand]\cite{SE-Handbook-Working-Group:2015:00}
An asset or object $l_i \in L$ that is operated on or consumed during the execution of a process.  They are the inputs and outputs of systems processes and ``move" through the system.  
\end{defn}
It is important to recognize the system resources are classified into three categories.  $R = M \cup B \cup H$, where $M$ is the set of transformation resources, $B$ is the set of independent buffers, and $H$ is the set of transportation processes.  Furthermore, the system buffers $B_S=M \cup B$ are introduced as well.   Fig. \ref{CH6:fig:HFGT:AA:BD} shows this classification as a SysML block diagram.  Similarly, the system processes are classified as well.  $P = P_\mu \cup P_{\bar{\eta}}$, where $P_\mu$ is the set of transformation processes, and $P_{\bar{\eta}} = P_\gamma \mbox{\Cross} P_\eta$ is the set of refined transportation processes, and where $\mbox{\Cross}$ is the Cartesian product. Fig. \ref{CH6:fig:HFGT:AA:AD} shows the flow of system processes as an activity diagram\cite{Schoonenberg:2018:ISC-BK04}.  

\vspace{-0.1in}
\vspace{-0.2in}

Finally, HFGT makes extensive use of the total number of degrees of freedom (or system capabilities) $DOF_S$.  
\begin{equation}\label{CH6:Eq:DOFS1}
DOF_S=\sigma({\cal E}_S)=\sum_w^{\sigma(P)}\sum_v^{\sigma(R)} A_S(w,v)
\end{equation}

\vspace{-0.2in}
\subsection{Hetero-functional Graph Theory: Incidence Tensor}\label{CH6:subsec:B:HFGTincidencetensor}
The second hetero-functional graph theory modeling construct is the hetero-functional incidence tensor $\widetilde{\cal M}_\rho$\cite{Farid:2020:ISC-JR06}.  It defines the structural relationship between the system capabilities ${\cal E}_S$, the system operands $L$, and the system buffers $B_S$.  

\vspace{-0.1in}
\begin{equation}\label{CH6:CH6:eq:projHFIT}
\widetilde{\cal M}_\rho=\widetilde{\cal M}_\rho^+-\widetilde{\cal M}_\rho^-
\end{equation}
\vspace{-0.2in}
\begin{defn}[The Negative 3$^{rd}$ Order Hetero-functional Incidence Tensor $\widetilde{\cal M}_\rho^-$]\cite{Farid:2020:ISC-JR06}\label{CH6:def:HFITneg}
The negative hetero-functional incidence tensor $\widetilde{\cal M_\rho}^- \in \{0,1\}^{\sigma(L)\times \sigma(B_S) \times \sigma({\cal E}_S)}$  is a third-order tensor whose element $\widetilde{\cal M}_\rho^{-}(i,y,\psi)=1$ when the system capability ${\epsilon}_\psi \in {\cal E}_S$ pulls operand $l_i \in L$ from buffer $b_{s_y} \in B_S$.
\end{defn} 
\begin{defn}[The Positive  3$^{rd}$ Order Hetero-functional Incidence Tensor $\widetilde{\cal M}_\rho^+$]\cite{Farid:2020:ISC-JR06}\label{CH6:def:HFITpos}
The positive hetero-functional incidence tensor $\widetilde{\cal M}_\rho^+ \in \{0,1\}^{\sigma(L)\times \sigma(B_S) \times \sigma({\cal E}_S)}$  is a third-order tensor whose element $\widetilde{\cal M}_\rho^{+}(i,y,\psi)=1$ when the system capability ${\epsilon}_\psi \in {\cal E}_S$ injects operand $l_i \in L$ into buffer $b_{s_y} \in B_S$.
\end{defn}
\noindent These definitions can be used directly to determine the non-zero elements of the respective incidence tensor.  Alternatively, Farid et. al. have provided a method for their calculation from more fundamental hetero-functional graph theory concepts\cite{Farid:2020:ISC-JR06}.  

The development of the hetero-functional network minimum cost flow optimization program requires the matricization (or ``flattening") of the hetero-functional incidence tensor into a hetero-functional incidence tensor where the operand (i.e. first), and the buffer (i.e. second) dimension are combined.  The matricization function ${\cal F}_M()$ is adopted from \cite{Farid:2020:ISC-JR06}.  
\begin{align} \label{CH6:eq:matricizeHFIT}
\widetilde{M}_\rho&={\cal F}_M\left(\widetilde{{\cal M}}_\rho,[1,2],[3]\right)\\ \label{CH6:eq:matricizeHFIT-}
\widetilde{M}_\rho^-&={\cal F}_M\left(\widetilde{{\cal M}}_\rho^-,[1,2],[3]\right)\\ \label{CH6:eq:matricizeHFIT+}
\widetilde{M}_\rho^+&={\cal F}_M\left(\widetilde{{\cal M}}_\rho^+,[1,2],[3]\right)
\end{align}

\subsection{Timed Petri nets}\label{CH6:subsec:B:timedPN}
As mentioned previously, timed Petri nets serve as an intermediate modeling vehicle that facilitates the transformation of a hetero-functional graph into a hetero-functional network minimum cost flow optimization program.  

\begin{defn}[Continuous Marked Place-Transition Net (Graph) 
\cite{Popova-Zeugmann:2013:00,Girault:2013:00}]\label{CH6:defn:Place-TransitionNet}
A bipartite directed graph represented as a 5-tuple ${\cal N} = \{ S, {\cal E}, \textbf{M}, W, Q \}$, where 
\begin{itemize}
\item ${\cal N}$ is the place-transition net.
\item $S$ is a finite set of places.
\item ${\cal E}$ is a finite set of (instantaneous) transitions, such that $B \cap {\cal E} = \emptyset$ and $S \cup {\cal E} \not{=} \emptyset$.
\item $\textbf{M} \subseteq (S\times {\cal E}) \cup ({\cal E} \times S)$ is a set of arcs of size $\sigma(\mathbf{M})$ from places to transitions and from transitions to places in the graph. Furthermore, defined are the associated incidence matrix $M = M^+ - M^-$ where the positive incidence matrix  has element $M^+(s,e) \in \{0,1\}$ and the negative incidence matrix has element  $M^-(s,e) \in \{0,1\}$ for all $(s,e) \in S \times {\cal E}$.
\item $W : \mathbf{M} \rightarrow \mathds{R}$, is the set of weights on the arcs.
\item $Q : S\cup{\cal E} \rightarrow \mathds{R}$ is the marking of the place-transition net states. 
\end{itemize}
The definition of the weights W and the markings Q over the real numbers gives the Petri net its continuous rather than discrete nature.  
\end{defn}
 \begin{defn}[Timed Place-Transition Net Dynamics \cite{Popova-Zeugmann:2013:00}]\label{CH6:defn:TPNDyn}
Given a binary input firing vector $U^-[k]$ and a binary output firing vector $U^+[k]$ both of size $\sigma({\cal E}) \times 1$, and the positive and negative components ${M}^+$ and ${M}^-$ of the Petri net incidence matrix of size $\sigma(S) \times \sigma ({\cal E})$, the evolution of the marking vector $Q \in \mathds{R}^{\sigma(S)+ \sigma({\cal E})}$ is given by the state transition function $\Phi_T(Q[k],U^-[k],U^+[k])$:
\begin{equation}\label{CH6:eq:Phi}
Q[k+1]=\Phi_T(Q[k],U^-[k], U^+[k])
\end{equation}
where $Q=[Q_{B}; Q_{\cal E}]$ and 
\begin{align}\label{CH6:CH6:eq:Q_B:TPN}
Q_{B}[k+1]=&Q_{B}[k]+{M}^+U^+[k]-{M}^-U^-[k]\\ \label{CH6:CH6:eq:Q_E:TPN}
Q_{\cal E}[k+1]=&Q_{\cal E}[k]-U^+[k] +U^-[k] \\ \label{CH6:eq:sync}
U_\psi^+[k+k_{d\psi}] =& U_\psi^-[k]
\end{align}
and where $U_\psi^-[k]$ indicates the $\psi^{th}$ element of the $U^-[k]$ vector and  Eq. \ref{CH6:eq:sync} allows for a transition duration of $k_{d\psi}$ between the negative and positive firing vectors.    
\end{defn}

\vspace{-0.2in}
\subsection{Hetero-functional Graph Theory:  Service Model} \label{CH6:subsec:B:HFGTservices}
The third hetero-functional graph theory modeling construct utilizes Defn. \ref{CH6:defn:Place-TransitionNet} and is called the service model.   It describes the collective behavior of operands in an engineering system.  It is composed of one service Petri net and one service feasibility matrix for each operand.  
\begin{defn}[Service Petri Net\cite{Farid:2014:ISC-C37,Farid:2014:ISC-C38,Farid:2015:ISC-J19,Khayal:2017:ISC-J35,Schoonenberg:2017:IEM-J34}]\label{CH6:defn:ServicePetriNet} Given service $l_i$, a service net ${\cal N}_{l_i}$ is marked place-transition net where 
\begin{equation}
	{\cal N}_{l_i} = \{S_{l_i}, {\cal E}_{l_i}, \textbf{M}_{l_i}, W_{l_i}, Q_{l_i}\}
\end{equation}
where
\begin{itemize}
\item $S_{l_i}$ is the set of places describing a set of service states.
\item ${\cal E}_{l_i}$ is the set of transitions describing service activities.
\item $\textbf{M}_{l_i} \subseteq (S_{l_i} \times {\cal E}_{l_i}) \cup ({\cal E}_{l_i} \times S_{l_i})$ is the set of arcs describing the relations of (service states to service activities) and (service activities to service states). Furthermore, defined are the associated incidence matrix $M_{l_i} = M^+_{l_i} - M^-_{l_i}$ where the positive incidence matrix has element $M^+_{l_i}(s_{\zeta l_i},e_{xl_i}) \in \{0,1\}$ and the negative incidence matrix has element  $M^-_{l_i}(s_{\zeta l_i},e_{xl_i}) \in \{0,1\}$ for all $(s_{\zeta l_i},e_{xl_i}) \in S_{l_i} \times {\cal E}_{l_i}$.
\item $W_{l_i} : \textbf{M}_{l_i} \rightarrow [0 \dots 1]$ is the set of weights on the arcs describing the service transition probabilities for the arcs.  
\item $Q_{l_i}$ is the Petri net marking representing the set of service states.
\end{itemize}
\vspace{-0.1in}
\end{defn}
\vspace{-0.2in}
\begin{figure}[h]
\centering
\includegraphics[width=3in]{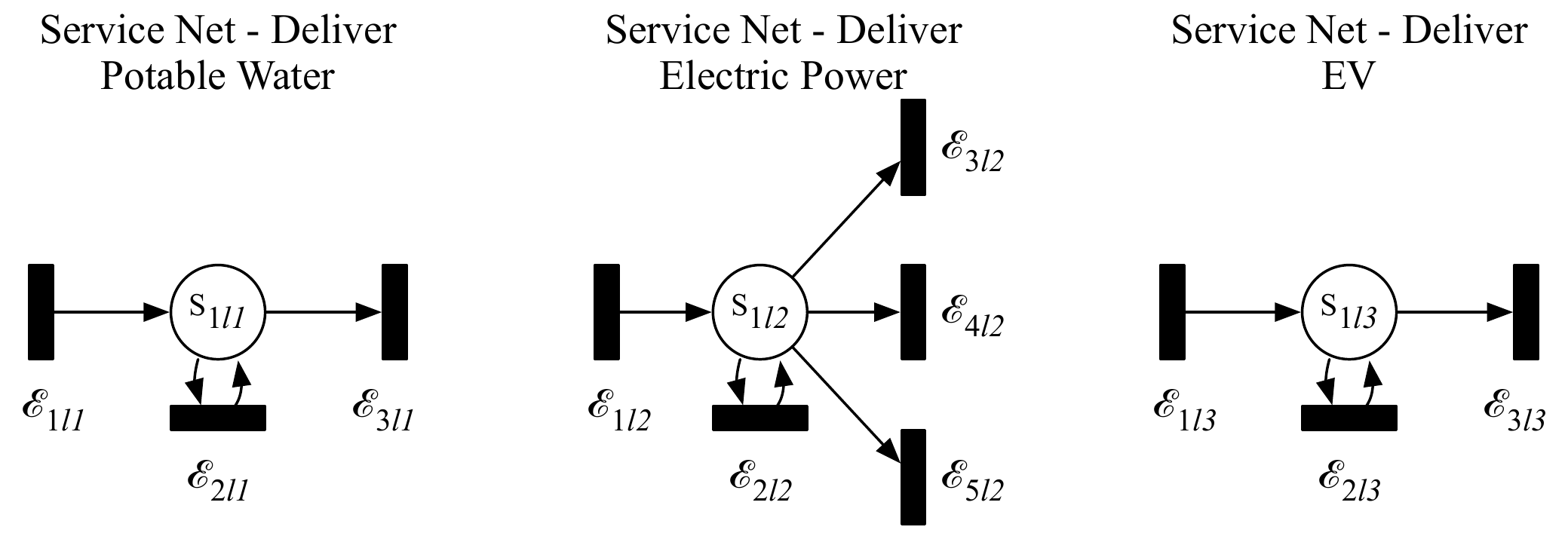}
\caption{Three service nets.  One for each operand (a) Water, (b) Power, and (c) Electric Vehicle\cite{Schoonenberg:2018:ISC-BK04}.}
\label{CH6:fig:HFGT:AA:SN}
\end{figure}
\vspace{-0.1in}
Fig. \ref{CH6:fig:HFGT:AA:SN} displays a service net for three operands.  The places track the operand state, and the transitions evolve the state of the operand. Furthermore, the transitions can ``create" or ``destroy" operands, by transitions that do not have an origin or destination respectively.

Service Petri nets have the following dynamics:
\begin{defn}[Service Net Dynamics \cite{Popova-Zeugmann:2013:00}]\label{CH6:defn:ServicePetriNetDyn}
Given a binary input firing vector $U^+_{l_i}[k]$ and a binary output firing vector $U^-_{l_i}[k]$ both of size $\sigma({\cal E}_{l_i}) \times 1$, and the positive and negative components ${M}^+_{l_i}$ and ${M}^-_{l_i}$ of the Petri net incidence matrix of size $\sigma(S_{l_i}) \times \sigma ({\cal E}_{l_i})$, the evolution of the marking vector $Q_{l_i}$ is given by the state transition function $\Phi_{l_i}(Q_{l_i}[k],U^-_{l_i}[k],U^+_{l_i}[k])$:
\begin{equation}\label{CH6:CH6:eq:2D:PhiL}
Q_{l_i}[k+1]=\Phi_{l_i}(Q_{l_i}[k],U_{l_i}^-[k], U_{l_i}^+[k])
\end{equation}
where $Q_{l_i}=[Q_{Sl_i}; Q_{{\cal E}l_i}]$ and 
\begin{align}\label{CH6:eq:Q_SL}
Q_{Sl_i}[k+1]=&Q_{Sl_i}[k]+{M}_{l_i}^+U_{l_i}^+[k]-{M}_{l_i}^-U_{l_i}^-[k]\\ \label{CH6:CH6:eq:Q_EL:TPN}
Q_{{\cal E}l_i}[k+1]=&Q_{{\cal E}l_i}[k]-U_{l_i}^+[k] +U_{l_i}^-[k]
\end{align}
The duration of the service net transitions is discussed specifically in Sec. \ref{CH6:subsec:HFGTdyn:services}.  
\end{defn}

In addition to the service petri net, the hetero-functional graph theory service model includes the service feasibility matrix.   
\begin{defn}[Service-Capability Feasibility Matrix\cite{Farid:2020:ISC-JR06}]\label{CH6:defn:CH4:SFM}
For a given service $l_i$, a binary matrix of size $\sigma({\cal E}_{l_i}) \times \sigma({\cal E}_S)$ whose value $\widetilde{\Lambda}_{i}(x,\psi)=1$ if $e_{xl_i}$ realizes capability $e_{s\psi}$. Furthermore:
\begin{equation}
\widetilde{\Lambda}_{i} = \widetilde{\Lambda}^+_{i} \oplus \widetilde{\Lambda}^-_{i}
\end{equation}
such that $\widetilde{\Lambda}^+_{i}$ realizes capability $e_{s\psi}$ to generate the output $l_i$ and $\widetilde{\Lambda}^-_{i}$ realizes capability $e_{s\psi}$ and uses $l_i$ as its input.
\end{defn}

\noindent The service feasibility matrix couples the operand behavior to the hetero-functional graph theory incidence tensor.  

\vspace{-0.1in}
\subsection{Multi-sets}\label{CH6:subsec:B:bag}
In order to discuss arc-constant colored Petri nets in the next subsection, a mathematical foundation for multi-sets is introduced here.  
\begin{defn}[Multi-set or Bag \cite{Jensen:1992:00}]\label{CH6:defn:bag}
A multi-set $m$, over a non-empty set ${\cal S}$, is a function of $m\in [{\cal S} \rightarrow \mathds{N}]$. The non-negative integer $m(s) \in \mathds{N}$ is the number of appearances of the element $s$ in the multi-set $m$. The multi-set $m$ is represented by a formal sum:
\begin{equation}\label{CH6:eq:multiset}
\sum_{s\in {\cal S}} m(s)'s
\end{equation}
${\cal S}_{MS}$ denotes the set of all multi-sets over ${\cal S}$. The non-negative integers $\{m(s)\ |\ s \in {\cal S}\}$ are called the coefficients of the multi-set $m$, and $m(s)$ is called the coefficient of $s$. An element $s \in {\cal S}$ is said to belong to the multi-set $m$ iff $m(s) \not{=} 0$, and thus $s \in m$. 
\end{defn}
\noindent In this work, this multi-set definition is relaxed so that $m(s) \in \mathds{R}^+$ to allow for fractional members of a set.  Finally, multi-sets admit arithmetic operations as expected.  
\begin{align}
    m_1 + m_2 &= \sum_{s\in {\cal S}} (m_1(s) + m_2(s))`s \\ \label{CH6:CH6:eq:multisetsubtraction}
    m_2 - m_1 &= \sum_{s\in {\cal S}} (m_2(s) - m_1(s))`s \\
    n * m &= \sum_{s \in {\cal S}} (n * m(s))`s
\end{align}
where $m_1, m_2 \in S_{MS}$ and all $n\in \mathds{R}^+$.   

\vspace{-0.1in}
\subsection{Arc-Constant Colored Petri Nets}\label{CH6:subsec:B:accpn}
In addition to timed place-transition nets, arc-constant colored Petri nets (ac-CPN) serve as an intermediate modeling vehicle that facilitates the transformation of a hetero-functional graph into a hetero-functional network minimum cost flow optimization program.  More specifically, ac-CPNs are used to introduce operand heterogeneity to the Petri net logic. 
\begin{defn}[Arc-constant colored Petri net (ac-CPN)\cite{Popova-Zeugmann:2013:00,Girault:2013:00}]\label{CH6:def:B:acCPN:acCPN}
An arc-constant colored Petri net  ${\cal N}_{\cal C}$ is defined by a tuple ${\cal N}_{\cal C} = \{S_C, {\cal E}_C, \textbf{M}_{\cal C}, {\cal C}, cd,Q_{\cal C}\}$, where
\begin{itemize}
\item $S_C$ is a finite set of places,
\item ${\cal E}_C$ is a finite set of transitions disjoint from $S_C$,
\item $\textbf{M}_{\cal C} \subseteq (S_C \times {\cal E}_C) \cup ({\cal E}_C \times S_C)$. The associated incidence matrix $M_{\cal C} = M^+_{\cal C} - M^-_{\cal C}$ where the positive incidence matrix $M^+_{\cal C} \in {\cal B}^{|S_C|\times |{\cal E_C}|}$ has element $M^+_{\cal C}(s_c,e_c) \in \text{Bag}(cd(s_c))$ and the negative incidence matrix $M^-_{\cal C} \in {\cal B}^{|S_C|\times |{\cal E}_C|}$ has element  $M^-_{\cal C}(s_c,e_c) \in \text{Bag}(cd(s_c))$ for all $(s_c,e_c) \in S_{\cal C} \times {\cal E}_{\cal C}$.
\item ${\cal C}$ is the set of color classes.
\item $cd: S_{\cal C} \rightarrow {\cal C}$ is the color domain mapping. 
\item $Q_{\cal C} \in \text{Bag}(cd(s_c))$ is the marking vector of the arc-constant Colored Petri Net states. It is equal in size to the number of places.  
\end{itemize}
Note that ${\cal B}=\mbox{Bag}(A)$, where $A$ is the union of all color sets ${\cal C}$. Furthermore, the difference operator in $M_{\cal C} = M^+_{\cal C} - M^-_{\cal C}$ follows Eq. \ref{CH6:CH6:eq:multisetsubtraction}.  Finally, in comparison to the Place-Transition Net, the arc weights of an ac-CPN are integrated into the incidence matrices directly and the marking of the net is now over $\text{Bag}(cd(s_c))$ instead of over the set of positive real numbers.
\end{defn}

\begin{defn}[Arc-Constant Colored Petri Net State Transition Function $\Phi_{\cal C}()$]\label{CH6:defn:B:acCPN:Phi}
\begin{equation}\label{CH6:eq:B:acCPN:Phi}
Q_{\cal C}[k+1]=\Phi_{\cal C}(Q_{\cal C}[k],U_{\cal C}^-[k], U_{\cal C}^+[k]) \quad \forall k \in \{1, \dots, K\}
\end{equation}
where $Q_{\cal C}=[Q_{B{\cal C}}; Q_{{\cal E C}}]$ and 
\begin{align}\label{CH6:eq:B:acCPN:Q_BC} 
Q_{B{\cal C}}[k+1]=&Q_{B{\cal C}}[k]&+{M}_{\cal C}^+&U_{\cal C}^+[k] &-{M}_{\cal C}^-&U_{\cal C}^-[k] \\\label{CH6:eq:B:acCPN:Q_EC} 
Q_{{\cal EC}}[k+1]=&Q_{{\cal EC}}[k]&-&U_{\cal C}^+[k] &+&U_{\cal C}^-[k] \\ \label{CH6:eq:B:acCPN:ProcessTime}
 U_{{\cal C}\psi}^+[k+k_{d\psi}]= & U_{{\cal C}\psi}^-[k]
\end{align}
$U_{{\cal C}\psi}^-[k]$ indicates the $\psi^{th}$ element of the $U_{\cal C}^-[k]$ vector and  Eq. \ref{CH6:eq:B:acCPN:ProcessTime} allows for a transition duration of $k_{d\psi}$ between the negative and positive firing vectors.    
\end{defn}

While ac-CPNs are valuable tool for modeling, verification, and visualization, they must be transformed into place-transition nets prior to their use in an optimization setting.  Jensen has defined the steps necessary for such a transformation\cite{Jensen:1992:00}; which is summarized here using a tensor-based treatment.  

\begin{alg}[Conversion from an ac-CPN to a PN]\label{CH6:CH6:alg:acCPN-CPN}
\textbf{Input:}  ${\cal N}_{\cal C} = \{S_C, {\cal E}_C, \textbf{M}_{\cal C}, {\cal C}, cd,Q_{\cal C}\}$\\
\textbf{Output:}  ${\cal N} = \{ S, {\cal E}, \textbf{M}, W, Q \}$
\begin{enumerate}
\item Split the places of the ac-CPN for each color set.  $S = {\cal C} \mbox{\Cross} S_C$.
\item Retain the transitions of the ac-CPN.  ${\cal E} = {\cal E}_C$.
\item Redefine the multi-set negative incidence matrix $M^-_{\cal C}$ as a third-order negative incidence tensor ${\cal M}^-_{\cal C}$ where ${\cal M}^-_{\cal C}(c,s_c,e_c)=M^-_{\cal C}(s_c,e_c)'c$.   Matricize this tensor along the first two dimensions.  $M^-={\cal F}_M\left({\cal M}_{\cal C}^-,[1,2],[3]\right)$.  
\item Redefine the multi-set positive incidence matrix $M^+_{\cal C}$ as a third-order negative incidence tensor ${\cal M}^+_{\cal C}$ where ${\cal M}^+_{\cal C}(c,s_c,e_c)=M^+_{\cal C}(s_c,e_c)'c$.   Matricize this tensor along the first two dimensions.  $M^+={\cal F}_M\left({\cal M}_{\cal C}^+,[1,2],[3]\right)$.  
\item Redefine the initial multi-set marking vector $Q_{B{\cal C}}[0]$ as a matrix ${\cal Q}_{B{\cal C}}[0]$  where ${\cal Q}_{B{\cal C}}(c,s_c)[0]= Q_{B{\cal C}}(s_c)'c[0]$.  The vectorize this matrix.  $Q_{B}[0]= vec({\cal Q}_{B{\cal C}}[0])$.  
\item Retain the initial conditions of the ac-CPN transitions.  $Q_{\cal E}[0]= {\cal Q}_{{\cal E}C}[0]$.  
\end{enumerate}\vspace{-0.2in}
\end{alg}

\vspace{-0.2in}
\section{Hetero-functional Network Dynamics}\label{CH6:sec:HFGTdynamics}
Given the foundation of hetero-functional graph theory and Petri-net definitions provided above, this paper now derives the Hetero-functional Network Dynamics. The dynamic model consists of three parts: (1) the Engineering System Net, which represents the dynamics of the engineering system, (2) the Service Net, which represents the dynamics of the system operands, and (3) the Synchronization Matrix, which couples the operand behavior to the engineering system net behavior.  The hetero-functional network dynamics are modeled in discrete time.  Continuous time dynamics may be discretized into discrete-time\cite{Ogata:1994:00} and discrete-event dynamics can be given a system clock and scheduled event list\cite{Cassandras:2007:01} to recover discrete-time dynamics.   The three parts of the hetero-functional network dynamics are now discussed in sequence. 

\vspace{-0.1in}
\subsection{Engineering System Net}\label{CH6:subsec:HFGTdyn:esn}
The engineering system net describes the dynamics of the engineering system.  
\begin{defn}[Engineering System Net]\label{CH6:def:HFGTdyn:esn:ESN}
An arc-constant colored Petri net ${\cal N}_{\cal C} = \{B_S, {\cal E}_S, \textbf{M}_{\cal C}, L, cd,Q\}$, where
\begin{itemize}
\item $B_S$ system buffers are the set of places,
\item ${\cal E}_S$ system capabilities are the set of transitions (disjoint from $B_S$),
\item $\textbf{M}_{\cal C} \subseteq (B_S\times {\cal E}_S) \cup ({\cal E}_S \times B_S) $. The associated incidence matrix $M_{\cal C} = M^+_{\cal C} - M^-_{\cal C}$ such that 
\begin{align}\label{CH6:eq:acCPNPre}
M^-_{\cal C}(y, \psi) = \sum_{l_i \in L} \left(\widetilde{\cal M}^-_\rho(l_i, y, \psi) \right)' l_i \quad \in \{l_1, \dots, l_{\sigma(L)}\}\\ \label{CH6:eq:acCPNPost}
M^+_{\cal C}(y, \psi) = \sum_{l_i \in L} \left(\widetilde{\cal M}^+_\rho(l_i, y, \psi) \right)' l_i \quad \in \{l_1, \dots, l_{\sigma(L)}\}
\end{align}\vspace{-0.1in}
\item $L$ (system operands) are the set of color classes.
\item $cd: B_S \rightarrow L$ is the color domain mapping. 
\item $Q \in \text{Bag}(cd(s))$ is the marking vector of the engineering system net.  It represents the state of the engineering system.  
\end{itemize}\vspace{-0.25in}
\end{defn}
\noindent Here, it is important to recognize that the positive and negative hetero-functional incidence tensors indicate the presence of ``colored" arcs in the arc-constant colored Petri net.  Consequently, the hetero-functional incidence tensor can be used to straightforwardly recover the engineering systems behavior via the arc-constant colored Petri net state transition function  $\Phi_{\cal C}()$ (Defn. \ref{CH6:defn:B:acCPN:Phi}).  Furthermore, from a physics perspective, the engineering system net as defined above imposes continuity laws for all colored-operands at all system buffers.  Finally, this engineering system definition provided is a generalization of the one used in prior hetero-functional graph theory work for transportation systems\cite{Baca:2013:ETS-C11,Baca:2013:ETS-C23,Viswanath:2013:ETS-J08}, electrified transportation systems\cite{Farid:2016:ETS-BC05,Farid:2016:ETS-J27,vanderWardt:2017:ETS-J33,Viswanath:2014:ETS-C33}, production systems\cite{Farid:2007:IEM-TP00,Farid:2008:IEM-J04,Farid:2008:IEM-J06,Farid:2015:ISC-J19,Farid:2017:IEM-J13,Farid:2007:IEM-J02,Farid:2008:IEM-J05,Farid:2015:IEM-J23}, and microgrid-enabled production systems\cite{Schoonenberg:2015:IEM-C48,Schoonenberg:2017:IEM-J34}.

\vspace{-0.1in}
\subsection{Device Model Refinement of the Engineering System Net}\label{CH6:subsec:HFGTdyn:esnDM}
In addition to the continuity laws imposed by the engineering system net defined in the previous section, a set of device models must be added to describe the behavior of each system capability (or degree of freedom).  The nature of the device model depends on 1.) the type of engineering system, 2.) the nature of each capability, and 3.) the resolution (or degree of decomposition) by which the capability has been defined.  In time-driven systems with engineering physics and ``elemental" capabilities, these device models are constitutive laws (e.g. Ohm's resistor law, the capacitor law, and the inductor law) and compatibility laws (e.g. Kirchoff's Voltage law for electrical circuits) \cite{Schavemaker:2008:00,Farid:2015:SPG-J17}.  In such cases, the structural degrees of freedom (i.e. system capabilities) are equivalent to the degrees of freedom (i.e. generalized coordinates) in engineering physics\cite{Farid:2007:IEM-TP00,Farid:2008:IEM-J05,Farid:2008:IEM-J04,Farid:2015:ISC-J19}.  In other cases (e.g. power systems engineering), many elemental capabilities are combined into a single capability with a complex device model expressed as a set of simultaneous differential algebraic equations\cite{Farid:2015:SPG-J17,Gomez-Exposito:2008:00}.  

Given the tremendous diversity of engineering system device models, for the purposes of the hetero-functional network minimum cost flow optimization, this work restricts itself to device models that create a fixed ratio between input and output operands ($L$) for each of the system process ($P$).  These ratios are most easily implemented in a positive and negative device model matrix.  
\begin{defn}[Positive Device Model Matrix ]\label{CH6:def:posDevModelMat}
A matrix $D^+_{R} \in \mathds{R}^{+ \sigma(L) \times \sigma(P)}$ whose element $D^+_{R}(i,w)$ describes the relative quantity of operand $l_i$ ejected by process $p_w$.   
\end{defn}
\begin{defn}[Negative Device Model Matrix]\label{CH6:def:negDevModelMat}
A matrix $D^-_{R} \in \mathds{R}^{+ \sigma(L) \times \sigma(P)}$ whose element $D^-_{R}(i,w)$ describes the relative quantity of operand $l_i$ consumed by process $p_w$.   
\end{defn}

The primary advantage of using device models of this form is that they can be readily folded into the positive and negative hetero-functional incidence tensors respectively.  
\vspace{-0.1in}
\begin{align}\label{CH6:CH6:eq:HFITwDMpos}
\widehat{\cal M}^+_\rho = \left( \mathds{1}^{\sigma(B_S)} \circ \left( \mathds{1}^{\sigma(R)T} \otimes D^+_R \right)\mathds{P}^T_S \right)^T \odot \widetilde{\cal M}^+_\rho \\ \label{CH6:CH6:eq:HFITwDMneg}
\widehat{\cal M}^-_\rho = \left( \mathds{1}^{\sigma(B_S)} \circ \left( \mathds{1}^{\sigma(R)T} \otimes D^+_R \right)\mathds{P}^T_S  \right)^T \odot \widetilde{\cal M}^-_\rho 
\end{align}
where $\circ$ is the third-order outer product\cite{Kolda:2009:00,Kolda:2006:00}, and $\widehat{\cal M}^+_\rho$ and $\widehat{\cal M}^-_\rho$ are the positive and negative third-order device model refined hetero-functional incidence tensors of size $\sigma(L)\times\sigma(B_S)\times\sigma({\cal E}_S)$. These refined hetero-functional incidence tensors are then reincorporated directly into engineering system net (in Defn. \ref{CH6:def:HFGTdyn:esn:ESN}).

\vspace{-0.15in}
\subsection{Operand Behavior with the Service Model}\label{CH6:subsec:HFGTdyn:services}
The second element in the hetero-functional network dynamics is  the system operand behavior through Service Nets (Defn. \ref{CH6:defn:ServicePetriNet}) and their dynamics (Defn. \ref{CH6:defn:ServicePetriNetDyn}).  These definitions are adopted directly into the hetero-functional network dynamics without change.

\vspace{-0.15in}
\subsection{Synchronization Matrix}\label{CH6:subsec:HFGTdyn:sync} 
In hetero-functional graph theory, the engineering system net and the service nets are coupled through the service feasibility matrices (Defn. \ref{CH6:defn:CH4:SFM}). The coupling of their dynamics is achieved through the synchronization of the engineering system net and service net firing vectors. The state of the engineering system net is distinct from the state of the service net, but the transitions of both nets are coupled in time. The negative firing vectors indicate the start of transitions, they are synchronized by the negative service feasibility matrix $\widetilde{\Lambda}^-_i$. The positive firing vectors indicate the end of transitions, they are synchronized by the positive service feasibility matrix $\widetilde{\Lambda}^+_i$. 

The service synchronization must, however, also reflect the device models as implemented in the engineering system net. Consequently, the service feasibility matrices are first converted to the Synchronization Matrices:
\begin{align}\label{CH6:CH6:eq:syncMatrixPos}
\widehat{\Lambda}^+_i = \widetilde{\Lambda}^+_i \odot &\left( \left[e^{\sigma(L)T}_{i} \mathds{P}_S(\mathds{1}^{\sigma(R)T} \otimes D^+_R) \right] \otimes \mathds{1}^{\sigma({\cal E}_{l_i})} \right)  \\  \label{CH6:CH6:eq:syncMatrixNeg} 
\widehat{\Lambda}^-_i = \widetilde{\Lambda}^-_i \odot &\left( \left[ e^{\sigma(L)T}_{i} \mathds{P}_S(\mathds{1}^{\sigma(R)T} \otimes D^-_R) \right] \otimes \mathds{1}^{\sigma({\cal E}_{l_i})} \right) \\\nonumber & \qquad \forall i \in \{1, \dots, \sigma(L)\}
\end{align}
Then, the positive and negative firing vectors of the engineering system net and service nets are synchronized through the service synchronization equations:
\begin{align}\label{CH6:CH6:eq:SyncPosMultiSet}
    U_{l_i}^+[k] = \widehat{\Lambda}_i^+ &U_{{\cal C}}^+[k]\quad \forall i \in \{1, \dots, \sigma(L)\},\ k \in \{1, \dots, K\} \\ \label{CH6:CH6:eq:SyncNegMultiSet}
    U_{l_i}^-[k] = \widehat{\Lambda}_i^- &U_{{\cal C}}^-[k]\quad \forall i \in \{1, \dots, \sigma(L)\},\ k \in \{1, \dots, K\}
\end{align}
Note that the duration of transitions in the service net is a result of the duration of transitions in the engineering system net.

\vspace{-0.1in}
\section{Hetero-functional Network Minimum Cost Flow}\label{CH6:sec:HFGTprogram} 
This section develops the hetero-functional network minimum cost flow optimization program so as to optimize the dynamic system model developed in the previous section (Sec. \ref{CH6:sec:HFGTdynamics}). The first four constraints incorporate the engineering system net (Sec. \ref{CH6:subsec:HFGTprog:esn}) and service net dynamics (Sec. \ref{CH6:subsec:HFGTprog:ssn}), their synchronization (Sec. \ref{CH6:subsec:HFGTprog:sync}), and their transition duration (Sec. \ref{CH6:subsec:HFGTprog:proctime}). The section then defines the boundary constraints (Sec. \ref{CH6:subsec:HFGTprog:demand}), the initial and final conditions (Sec. \ref{CH6:subsec:HFGTprog:ICFC}), the capacity constraints (Sec. \ref{CH6:subsec:HFGTprog:capacity}), and the objective function (Sec. \ref{CH6:subsec:HFGTprog:objfun}). Finally, Sec. \ref{CH6:subsec:HFGTprog:full} provides the compiled optimization program.

\vspace{-0.1in}
\subsection{Engineering System Net}\label{CH6:subsec:HFGTprog:esn}
The engineering system net was defined as an ac-CPN in Sec. \ref{CH6:subsec:HFGTdyn:esn}. The state of the ac-CPN is defined as a multiset, which cannot be optimized with a conventional quadratic program over reals. It is therefore necessary to convert the ac-CPN to a regular Petri net using Algorithm \ref{CH6:CH6:alg:acCPN-CPN}. 

As a result of the conversion, the engineering system dynamics are now described by a net with the following properties:
\begin{itemize}
    \item $S$ is the set of places with length: $\sigma(L)\sigma(B_S)$,
    \item ${\cal E}$ is the set of transitions with length: $\sigma({\cal E}_S)$,
    \item $\textbf{M}$ is the set of arcs, with the associated incidence matrices: $M = M^+ - M^-$,
    \item $W$ is the set of weights on the arcs, as captured in the incidence matrices,
    \item $Q$ is the marking vector for both the set of places and the set of transitions. 
\end{itemize}
The state transition equations of the engineering system net are:
\begin{equation}\label{CH6:eq:PhiCPN}
Q[k+1]=\Phi_T(Q[k],U^-[k], U^+[k]) \quad \forall k \in \{1, \dots, K\}
\end{equation}
where $Q=[Q_{B}; Q_{\cal E}]$ and 
\begin{align}\label{CH6:CH6:eq:Q_B:HFNMCFprogram}
Q_{B}[k+1]&=Q_{B}[k]+{M}^+U^+[k]-{M}^-U^-[k] \\ \label{CH6:CH6:eq:Q_E:HFNMCFprogram}
Q_{\cal E}[k+1]&=Q_{\cal E}[k]-U^+[k] +U^-[k]
\end{align}
where $U^+ = U^+_{\cal C}$, $U^- = U^-_{\cal C}$, $Q_B$ has size $\sigma(L)\sigma(B_S) \times 1$, and $Q_{\cal E}$ has size $\sigma({\cal E}_S)\times 1$. These state transition functions are incorporated directly into the quadratic program in Sec. \ref{CH6:subsec:HFGTprog:full}. 

\vspace{-0.15in}
\subsection{Service Net}\label{CH6:subsec:HFGTprog:ssn}
The service net was defined as a Petri net in Sec. \ref{CH6:subsec:B:HFGTservices}. Recall that its dynamics are described by the transition function in Eq. \ref{CH6:CH6:eq:2D:PhiL}. The optimization program constraints require the concatenation of the state space equations over all the operands in the system:
$\Phi_L(Q_{L}[k],U^-_{L}[k],U^+_{L}[k])$, where $Q_{L}=[Q_{SL}; Q_{{\cal E}L}]$:
\begin{align}\label{CH6:CH6:eq:4:Q_SL}
Q_{SL}[k+1]&=Q_{SL}[k]+{M}_{L}^+U_{L}^+[k]-{M}_{L}^-U_{L}^-[k] \\ \label{CH6:CH6:eq:4:Q_EL}
Q_{{\cal E}L}[k+1]&=Q_{{\cal E}L}[k]-U_{L}^+[k] +U_{L}^-[k] \quad \forall k \in \{1, \dots, K\} 
\end{align}
where: $Q_{SL}$ has length $\sigma(Q_{SL}) = \sum_{l_i \in L} \sigma(S_{l_i})$ and is the vertical concatenation of the service net place markings for all operands in $L$:
\begin{equation}
Q_{SL} = 
\begin{bmatrix} 
Q_{Sl_1};\ \dots\ ; Q_{Sl_{\sigma(L)}} 
\end{bmatrix}
\end{equation}
$Q_{{\cal E}L}$ has length $\sigma(Q_{{\cal E}L}) = \sum_{l_i \in L} \sigma({\cal E}_{l_i})$ and is the vertical concatenation of the service net transition markings for all operands in $L$:
\vspace{-0.1in}
\begin{equation}
Q_{{\cal E}L} = 
\begin{bmatrix} 
Q_{{\cal E}l_1};\ \dots\ ; Q_{{\cal E}l_{\sigma(L)}}
\end{bmatrix}
\end{equation}
$U_{L}^+$ and $U_{L}^+$ are the vertical concatenations of the service net positive and negative firing vectors for all operands in $L$:
	\begin{equation}
	U^+_{L} = \begin{bmatrix} U^+_{l_1} \\ \vdots \\ U^+_{l_{\sigma(L)}} \end{bmatrix}, \qquad 
	U^-_{L} = \begin{bmatrix} U^-_{l_1} \\ \vdots \\ U^-_{l_{\sigma(L)}} \end{bmatrix}
	\end{equation}
	where $U^+_{L}$ and $U^-_{L}$ have size $\sigma(Q_{{\cal E}L}) \times 1$. 
Finally $M^+_L$ and $M^-_L$ are the block-diagonal positive and negative system service net incidence matrices:
\begin{equation}
M^+_L = \begin{bmatrix} M^+_{l_1} & \dots & 0 \\ \vdots & \ddots & \vdots \\ 0 & \dots & M^+_{l_{\sigma(L)}}\end{bmatrix},\ 
M^-_L = \begin{bmatrix} M^-_{l_1} & \dots & 0 \\ \vdots & \ddots & \vdots \\ 0 & \dots & M^-_{l_{\sigma(L)}}\end{bmatrix}
\end{equation}
where $M^+_L$ and $M^+_L$ have size $\sigma(Q_{SL}) \times \sigma(Q_{{\cal E}L})$.

\vspace{-0.1in}
\subsection{Synchronization Constraint}\label{CH6:subsec:HFGTprog:sync}
The synchronization of the engineering system net and the service nets was defined in Sec. \ref{CH6:subsec:HFGTdyn:sync}. The conversion from the engineering system firing vector $U_{\cal C}$ to the Petri net firing vector $U$ requires the conversion of Eqs.  \ref{CH6:CH6:eq:SyncPosMultiSet} and \ref{CH6:CH6:eq:SyncNegMultiSet} to:
\begin{align}
U^+_{L}[k] &= \widehat{\Lambda}^+ U^+[k] \qquad \forall k \in \{1, \dots, K\}\\
U^-_{L}[k] &= \widehat{\Lambda}^- U^-[k] \qquad \forall k \in \{1, \dots, K\}
\end{align}
where:
\vspace{-0.1in}
\begin{align}
\widehat{\Lambda}^+ = \begin{bmatrix}
\widehat{\Lambda}^+_1 \\ \vdots \\ \widehat{\Lambda}^+_{\sigma(L)}
\end{bmatrix}, \quad 
\widehat{\Lambda}^- = \begin{bmatrix}
\widehat{\Lambda}^-_1 \\ \vdots \\ \widehat{\Lambda}^-_{\sigma(L)}
\end{bmatrix}
\end{align}

\vspace{-0.1in}
\subsection{Duration Constraints}\label{CH6:subsec:HFGTprog:proctime}
The duration constraints are adopted from Eq. \ref{CH6:eq:B:acCPN:ProcessTime}. As the Engineering System Net firing vector is converted to a Petri net firing vector, the equation is defined as:
\begin{equation}\label{CH6:eq:durationequationESN}
U_\psi^+[k+k_{d\psi}] = U_\psi^-[k] \qquad \forall k \in \{1, \dots, K\}
\end{equation}
where $U_\psi^-[k]$ indicates the $\psi^{th}$ element of the $U^-[k]$ vector and where $k_{d\psi}$ is the duration of engineering system net transition $\psi$. 

\vspace{-0.1in}
\subsection{Boundary Constraints}\label{CH6:subsec:HFGTprog:demand}
The boundary constraints are the fifth element in the program. They define the interaction between the dynamic system and the context. These constraints are specifically used when modeling an \emph{open} system. The boundary constraints consist of two types: 1) demand constraints that control output transitions and 2) supply constraints that control input transitions. The demand constraints are imposed on $U^-[k]$:
\begin{align}
    D_{Bn}U^-[k] = C_{Bn}[k] \quad \forall k \in \{1, \dots, K \}
\end{align}
where $D_{Bn}$ is a transition selector matrix of size: $\sigma({\cal E}_{\text{Out}}) \times\ \sigma({\cal E}_S)$, with one filled element per row in the column of the selected transition, where $\sigma({\cal E}_{\text{Out}})$ is the \emph{number of output transitions}. Vector $C_{\text{demand}}[k]$ contains the demand data for each time step $k$. 

The supply constraints are imposed on $U^+[k]$:
\begin{align}
    D_{Bp}U^+[k] = C_{Bp}[k] \quad \forall k \in \{1, \dots, K \}
\end{align}
where $D_{Bp}$ is a transition selector matrix of size: $\sigma({\cal E}_{\text{In}}) \times\ \sigma({\cal E}_S)$, with one filled element per row in the column of the selected transition, where $\sigma({\cal E}_{\text{In}})$ is the \emph{number of input transitions}. Vector $C_{\text{supply}}[k]$ contains the supply data for each time step $k$. The boundary constraints are combined in a single equation:
\begin{equation}\label{CH6:eq:HFGTprog:boundary}
    \begin{bmatrix}
    D_{Bp} & \mathbf{0} \\ \mathbf{0} & D_{Bn}
    \end{bmatrix} \begin{bmatrix}
    U^+ \\ U^-
    \end{bmatrix}[k] = \begin{bmatrix}
    C_{Bp} \\ C_{Bn}
    \end{bmatrix}[k] \qquad \forall k \in \{1, \dots, K\}
\end{equation}

\subsection{Initial and Final Conditions}\label{CH6:subsec:HFGTprog:ICFC}
The initial conditions constrain the system at the initial time step: $k = 1$.  This allows the program to be used with a pre-populated system (also called a ``hot-start"). The initial conditions of the \emph{input} transitions should be left undetermined when modeling an \emph{open} system -- the optimization program will determine the quantities of the operands that need to enter the system in order to satisfy the demand. The initial condition constraints are:
\begin{align}
    \begin{bmatrix} Q_B ; Q_{\cal E} ; Q_{SL} \end{bmatrix}[k=1] = \begin{bmatrix} C_{B1} ; C_{{\cal E}1} ; C_{{SL}1} \end{bmatrix}
\end{align}
where \emph{``;"} is the MATLAB operator to define a vertically concatenated matrix.

The final conditions constrain the system at the final time step: $k = K+1$. The final conditions of the \emph{output} transitions should be left open when modeling an \emph{open} system. The state of those transitions in the last time step contains the cumulative outputs of that specific transition. Finally, in order to ensure that all tokens are accounted for, the negative firing vectors of the engineering system net and the system service net are set to zero. 
\begin{align}\nonumber
    \begin{bmatrix} Q_B;Q_{\cal E} ; Q_{SL} ; U^- ; U_L^- \end{bmatrix}&[k=K+1] =\\ &\begin{bmatrix} C_{BK} ; C_{{\cal E}K} ; C_{{SL}K} ; \mathbf{0} ; \mathbf{0} \end{bmatrix}
\end{align}

\vspace{-0.1in}
\subsection{Capacity Constraints}\label{CH6:subsec:HFGTprog:capacity}
The capacity constraints impose limits on the engineering system net. The capacity constraints limit the amount of each operand that can be fired at any point in time:
\begin{equation}
    U^-[k] \leq C_U \qquad \forall k \in \{1, \dots, K\}
\end{equation}

This equation is modified to account for system input transitions: transitions that input operands to the system without a predetermined value. These transitions are constrained specifically on the positive firing vectors. 
\begin{equation}
    \begin{bmatrix}
        D_{Cp} & \mathbf{0}\\
        \mathbf{0} & I^{\sigma({\cal E}_S)}
    \end{bmatrix} \begin{bmatrix}
    U^+ \\ U^-
    \end{bmatrix}[k] \leq C_U  \qquad \forall k \in \{1, \dots, K+1\}
\end{equation}
where $D_{Cp}$ selects the system input transitions without a predetermined value.

\subsection{Objective Function}\label{CH6:subsec:HFGTprog:objfun}
Finally, the objective function motivates the objective of the optimization program. It contains the cost or benefit of the execution of the decision variables. For the hetero-functional network minimum cost flow program, the cost is related to the execution of engineering system net transitions. However, when desired, cost can be imposed on other elements of the set of decision variables. The set of decision variables (as defined piece-wise in the previous sections) is defined as:
\begin{align}\nonumber
x[k] = &\begin{bmatrix} Q_B ; Q_{\cal E} ; Q_{SL} ; Q_{{\cal E}L} ; U^+ ; U^- ; U^+_L ; U^-_L \end{bmatrix}[k] \\ &\qquad\ \qquad\ \qquad\ \qquad\ \qquad \forall k \in \{1, \dots, K+1\}
\end{align}
where the size of the set of decision variables is:
\begin{equation}\label{CH6:eq:HFGTprog:sigmaX}
\sigma(x) = (K+1)\left( \sigma(B_S) + 3\sigma({\cal E}_S) + \sigma(Q_{SL}) + 3 \sigma(Q_{{\cal E}L}) \right)
\end{equation}

The cost function is imposed on the decision variables as either a linear or a quadratic function. This work introduces a quadratic objective function. The resulting objective function has the following form:
\begin{equation}
\text{minimize } Z = x^T F_{QP} x + f_{QP}^T x
\end{equation}
where $F_{QP} \geq 0$ is the quadratic cost coefficient (a matrix of size $\sigma(x) \times \sigma(x)$), and where $f_{QP} \geq 0$ is the linear cost coefficient (a vector of size $\sigma(x) \times 1$). Note that the quadratic cost matrix $F_{QP}$ is assumed to be diagonal. Furthermore, for all zero-valued elements on the diagonal, an infinitesimally small value may be added to ensure that the quadratic cost matrix is positive definite ($F_{QP} \succ 0$). This guarantees convexity of the quadratic program.

\subsection{Optimization Program Compilation}\label{CH6:subsec:HFGTprog:full}
Finally, this section compiles the elements of the optimization program to define the hetero-functional network minimum cost flow program. The canonical form of a linearly constrained quadratic program is presented below:
\begin{align}\label{ch6:eq:QPcanonicalform:1}
\text{minimize } Z &= x^T F_{QP} x + f_{QP}^T x \\ \label{ch6:eq:QPcanonicalform:2}
\text{s.t. } A_{QP}x &= B_{QP} \\ \label{ch6:eq:QPcanonicalform:3}
D_{QP}x &\leq E_{QP} \\ \label{ch6:eq:QPcanonicalform:4}
x &\geq 0, \quad x \in \mathds{R}
\end{align}
where:
\begin{itemize}
    \item $x$ has size $\sigma(x) \times 1$, as defined in Eq. \ref{CH6:eq:HFGTprog:sigmaX},
    \item $F_{QP}$ has size: $\sigma(x) \times \sigma(x)$,
    \item $f_{QP}$ has size: $\sigma(x) \times 1$,
    \item $A_{QP}$ has size: $\sigma(A_{QP}) \times \sigma(x)$
    \item $B_{QP}$ has size: $\sigma(A_{QP}) \times 1$,
    \item $D_{QP}$ has size: $\sigma(D_{QP}) \times \sigma(x)$
    \item $E_{QP}$ has size: $\sigma(D_{QP}) \times 1$.
\end{itemize}

Matrix $A_{QP}$ and vector $B_{QP}$ are constructed by concatenating eight constraints (Eqs.  \ref{CH6:eq:HFGTprog:comp:QB} through \ref{CH6:eq:HFGTprog:comp:Bound}) over all time steps $K$ with the initial and final condition constraints (Eqs.  \ref{CH6:eq:HFGTprog:comp:Init} and \ref{CH6:eq:HFGTprog:comp:Fini}):

\begin{align}\label{CH6:eq:HFGTprog:comp:QB}
-Q_{B}[k+1]+Q_{B}[k]+{M}^+U^+[k]-{M}^-U^-[k]=&0 \\  \label{CH6:eq:HFGTprog:comp:QE}
-Q_{\cal E}[k+1]+Q_{\cal E}[k]-U^+[k] +U^-[k]=&0 \\ \label{CH6:eq:HFGTprog:comp:PT}
 - U^+[k+k_{d\psi}]+ U^-[k] = &0 \\ \label{CH6:eq:HFGTprog:comp:QSL}
-Q_{SL}[k+1]+Q_{SL}[k]+{M}_{L}^+U_{L}^+[k]-{M}_{L}^-U_{L}^-[k]=&0 \\ \label{CH6:eq:HFGTprog:comp:QEL}
-Q_{{\cal E}L}[k+1]+Q_{{\cal E}L}[k]-U_{L}^+[k] +U_{L}^-[k]=&0 \\ \label{CH6:eq:HFGTprog:comp:Spos}
U^+_L[k] - \widehat{\Lambda}^+ U^+[k] =&0 \\ \label{CH6:eq:HFGTprog:comp:Sneg}
U^-_L[k] - \widehat{\Lambda}^- U^-[k] =&0 \\ \label{CH6:eq:HFGTprog:comp:Bound}
\begin{bmatrix}
    D_{Bp} & \mathbf{0} \\ \mathbf{0} & D_{Bn}
    \end{bmatrix} \begin{bmatrix}
    U^+ \\ U^-
    \end{bmatrix}[k] = \begin{bmatrix}
    C_{Bp} \\ C_{Bn}
    \end{bmatrix}&[k]
\end{align}
where Eqs. \ref{CH6:eq:HFGTprog:comp:QB} through \ref{CH6:eq:HFGTprog:comp:Bound} defined for all $k \in \{1, \dots, K\}$. The initial and final condition constraints are:
\begin{align}\label{CH6:eq:HFGTprog:comp:Init} 
\begin{bmatrix} Q_B ; Q_{\cal E} ; Q_{SL} \end{bmatrix}&[k=1] = \begin{bmatrix} C_{B1} ; C_{{\cal E}1} ; C_{{SL}1} \end{bmatrix} \\ \nonumber
\begin{bmatrix} Q_B ; Q_{\cal E} ; Q_{SL} ; U^- ; U_L^- \end{bmatrix}&[k=K+1] = \\ \label{CH6:eq:HFGTprog:comp:Fini} &\begin{bmatrix} C_{BK} ; C_{{\cal E}K} ; C_{{SL}K} ; \mathbf{0} ; \mathbf{0} \end{bmatrix}
\end{align}
Consequently, the number of rows in the $A_{QP}$ matrix is defined as:
\begin{align}\nonumber
    \sigma(A_{QP}) &= K\big[\sigma(Q_B) + 2\sigma(Q_{\cal E}) + \sigma(Q_{SL}) + 3\sigma(Q_{{\cal E}L})\ +\big. \\ \nonumber
    &\big. \sigma({\cal E}_{\text{Out}}) + \sigma({\cal E}_{\text{In}})\big]\ +  
    \sigma(Q_B) + \sigma(Q_{\cal E}) + \sigma(Q_{SL})\ + \\ 
    & \sigma(Q_B) + 2\sigma(Q_{\cal E}) + \sigma(Q_{SL}) + \sigma(Q_{{\cal E}L})
\end{align}
Note that the number of decision variables is defined over $K+1$ time steps to accommodate the mathematical structure of the state transition equations. 

The inequality constraints, $Dx \leq E$, contain the capacity constraints:
\begin{align}
    \begin{bmatrix}
        D_{Cp} & \mathbf{0}\\
        \mathbf{0} & I^{\sigma({\cal E}_S)}
    \end{bmatrix} \begin{bmatrix}
    U^+ \\ U^-
    \end{bmatrix}[k] \leq C_U  \qquad \forall k \in \{1, \dots, K+1\}
\end{align}
which is defined over the time steps $K+1$ to maintain consistency with the number of decision variables. The number of rows of the inequality matrix $D_{QP}$ is defined as:
\vspace{-0.1in}
\begin{align}
    \sigma(D_{QP}) = (K+1)\big[\sigma({\cal E}_{\text{In}}) + \sigma(Q_{\cal E}) \big]
\end{align}

\vspace{-0.1in}
\section{Illustrative Example: Hydrogen-Natural Gas System}\label{CH6:sec:IllustrativeExample}
This section introduces a test case to demonstrate the application of the hetero-functional network minimum cost-flow program. The section first introduces the context of the test case, then it provides the test case data and finally, the it introduces four optimization scenarios.
\vspace{-0.1in}
\subsection{Introduction}\label{CH6:subsec:IE:intro}
Test cases enable the study of modeling, simulation, and optimization methods of complex critical (infrastructure) systems \cite{Subcommittee:1979:00,Center-for-Water-Systems:2006:00,Farid:2015:ETS-J25}. The test case in this work is the first hydrogen-natural gas infrastructure test case to the knowledge of the authors. The test case is inspired by the Dutch natural gas system and the plans for a European hydrogen pipeline network \cite{Wang:2020:00} and it does not aim to represent the current or future system. 

The plans to develop hydrogen infrastructure are driven by the need for the reduction of carbon emissions. Electrolysis enables carbon-free generation of hydrogen from electric power and water. Consequently, hydrogen may serve as an intermediate mode of energy storage. A secondary benefit is that some industrial processes require a high-heat energy source. This is challenging to achieve through electric power, but hydrogen provides a (still expensive) alternative to natural gas and coal. Finally, natural gas is currently used as the energy source for the production of hydrogen. As a consequence, the hydrogen and natural gas system have interdependencies and overlap of their services. This interdependent system is especially challenging to operate and optimize.

\subsection{Test Case Data}\label{CH6:subsec:IE:data}

\begin{figure}[h]
\centering
\includegraphics[width=3in]{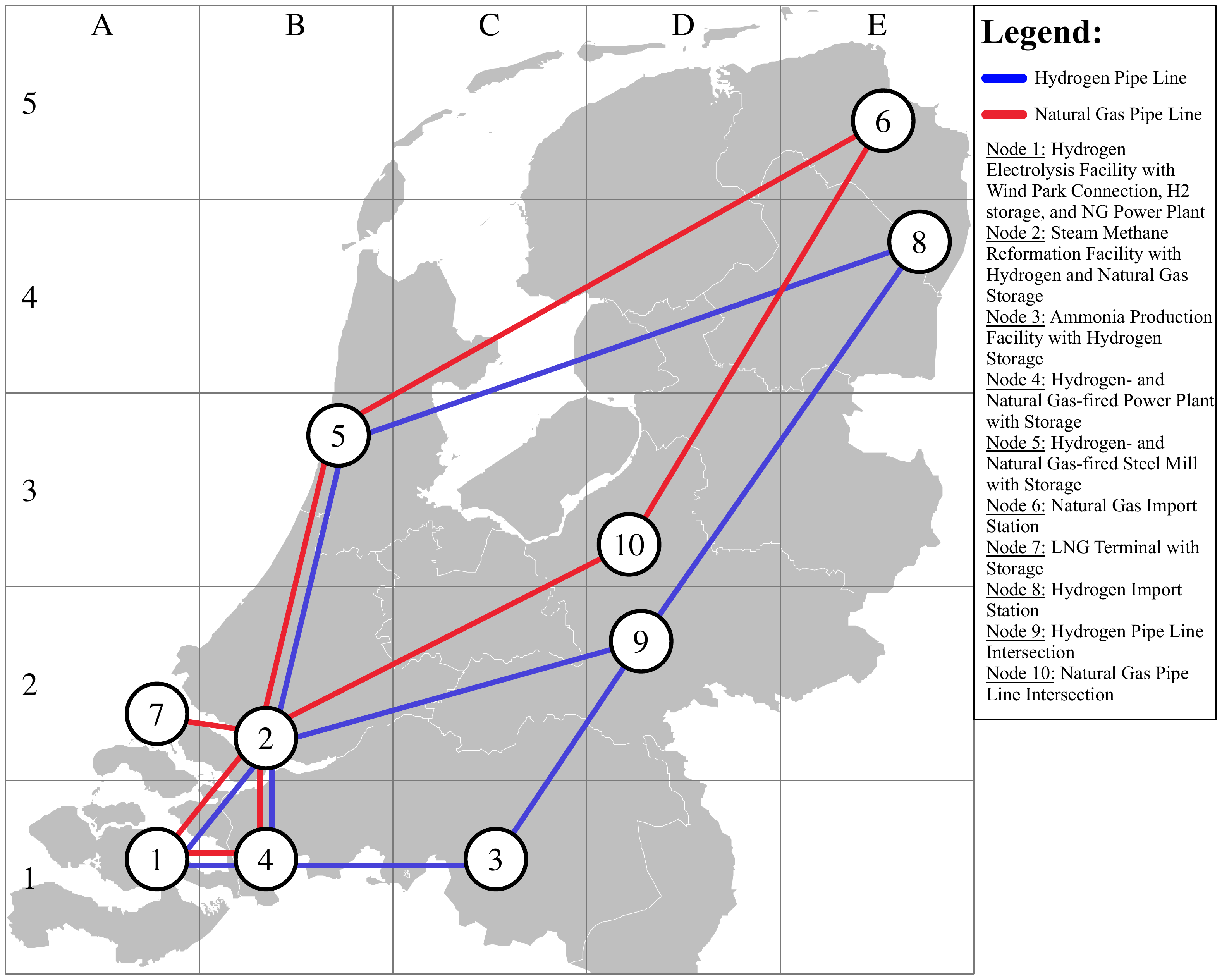}
\caption{Hydrogen Natural Gas Test Case. }\label{CH6:fig:testcaseNetwork}
\end{figure}

\begin{table*}[]
\tiny
\centering
\caption{Overview of the test case resources, processes, cost, capacity, and processing time.}\label{CH6:CH6:tab:testcaseResources}
\begin{tabular}{cllcclc}
\hline
\multicolumn{1}{|l|}{\textbf{Node \#}} & \multicolumn{1}{l|}{\textbf{Node Name}} & \multicolumn{1}{l|}{\textbf{Processes}} & \multicolumn{1}{l|}{\textbf{Quadratic Cost}} & \multicolumn{1}{l|}{\textbf{Linear Cost}} & \multicolumn{1}{l|}{\textbf{Capacity}} & \multicolumn{1}{l|}{\textbf{Processing Time}} \\ \hline
\multicolumn{1}{|c|}{\multirow{11}{*}{1}} & \multicolumn{1}{l|}{\multirow{11}{*}{\begin{tabular}[c]{@{}l@{}}Hydrogen \\ Electrolysis \\ Facility\end{tabular}}} & \multicolumn{1}{l|}{\begin{tabular}[c]{@{}l@{}}Electrolyze Water to\\ Hydrogen and Oxygen\end{tabular}} & \multicolumn{1}{c|}{-} & \multicolumn{1}{c|}{\$1000 / ton H$_2$} & \multicolumn{1}{l|}{3,000 ton H$_2$ / day} & \multicolumn{1}{c|}{2 days} \\ \cline{3-7} 
\multicolumn{1}{|c|}{} & \multicolumn{1}{l|}{} & \multicolumn{1}{l|}{\begin{tabular}[c]{@{}l@{}}Burn Natural Gas to\\ Generate Electric Power\end{tabular}} & \multicolumn{1}{c|}{0.01 $\$^2$/ton CH$_4$} & \multicolumn{1}{c|}{\$145 / ton CH$_4$} & \multicolumn{1}{l|}{3,000 ton CH$_4$ / day} & \multicolumn{1}{c|}{1 day} \\ \cline{3-7} 
\multicolumn{1}{|c|}{} & \multicolumn{1}{l|}{} & \multicolumn{1}{l|}{Import Electric Power} & \multicolumn{1}{c|}{-} & \multicolumn{1}{c|}{\$10 MWh} & \multicolumn{1}{l|}{100,000 MWh / day} & \multicolumn{1}{c|}{0 days} \\ \cline{3-7} 
\multicolumn{1}{|c|}{} & \multicolumn{1}{l|}{} & \multicolumn{1}{l|}{Import Water} & \multicolumn{1}{c|}{-} & \multicolumn{1}{c|}{-} & \multicolumn{1}{l|}{30,000 ton H$_2$O / day} & \multicolumn{1}{c|}{0 days} \\ \cline{3-7} 
\multicolumn{1}{|c|}{} & \multicolumn{1}{l|}{} & \multicolumn{1}{l|}{Export Water} & \multicolumn{1}{c|}{-} & \multicolumn{1}{c|}{-} & \multicolumn{1}{l|}{30,000 ton H$_2$O / day} & \multicolumn{1}{c|}{0 days} \\ \cline{3-7} 
\multicolumn{1}{|c|}{} & \multicolumn{1}{l|}{} & \multicolumn{1}{l|}{Import Oxygen} & \multicolumn{1}{c|}{-} & \multicolumn{1}{c|}{-} & \multicolumn{1}{l|}{30,000 ton O$_2$ / day} & \multicolumn{1}{c|}{0 days} \\ \cline{3-7} 
\multicolumn{1}{|c|}{} & \multicolumn{1}{l|}{} & \multicolumn{1}{l|}{Export Oxygen} & \multicolumn{1}{c|}{-} & \multicolumn{1}{c|}{-} & \multicolumn{1}{l|}{30,000 ton O$_2$ / day} & \multicolumn{1}{c|}{0 days} \\ \cline{3-7} 
\multicolumn{1}{|c|}{} & \multicolumn{1}{l|}{} & \multicolumn{1}{l|}{Export CO2} & \multicolumn{1}{c|}{-} & \multicolumn{1}{c|}{See Scenarios} & \multicolumn{1}{l|}{30,000 ton CO$_2$ / day} & \multicolumn{1}{c|}{0 days} \\ \cline{3-7} 
\multicolumn{1}{|c|}{} & \multicolumn{1}{l|}{} & \multicolumn{1}{l|}{Export Heat Loss} & \multicolumn{1}{c|}{-} & \multicolumn{1}{c|}{-} & \multicolumn{1}{l|}{30,000 MMBTU / day} & \multicolumn{1}{c|}{0 days} \\ \cline{3-7} 
\multicolumn{1}{|c|}{} & \multicolumn{1}{l|}{} & \multicolumn{1}{l|}{Store Hydrogen} & \multicolumn{1}{c|}{-} & \multicolumn{1}{c|}{\$ 0.1 / ton H$_2$} & \multicolumn{1}{l|}{21,000 ton H$_2$ / day} & \multicolumn{1}{c|}{1 day} \\ \cline{3-7} 
\multicolumn{1}{|c|}{} & \multicolumn{1}{l|}{} & \multicolumn{1}{l|}{Store Natural Gas} & \multicolumn{1}{c|}{-} & \multicolumn{1}{c|}{\$ 0.1 / ton CH$_4$} & \multicolumn{1}{l|}{100,000 ton CH$_4$ / day} & \multicolumn{1}{c|}{1 day} \\ \hline
\multicolumn{7}{c}{} \\ \hline
\multicolumn{1}{|c|}{\multirow{8}{*}{2}} & \multicolumn{1}{l|}{\multirow{8}{*}{\begin{tabular}[c]{@{}l@{}}Steam Methane \\ Reformation \\ Facility\end{tabular}}} & \multicolumn{1}{l|}{\begin{tabular}[c]{@{}l@{}}Reform Steam and\\ Methane to Hydrogen\\ and CO2\end{tabular}} & \multicolumn{1}{c|}{-} & \multicolumn{1}{c|}{\$ 1000 / ton H$_2$} & \multicolumn{1}{l|}{3,000 ton H$_2$ / day} & \multicolumn{1}{c|}{2 days} \\ \cline{3-7} 
\multicolumn{1}{|c|}{} & \multicolumn{1}{l|}{} & \multicolumn{1}{l|}{\begin{tabular}[c]{@{}l@{}}Burn Natural Gas to \\ Generate Industrial Heat\end{tabular}} & \multicolumn{1}{c|}{-} & \multicolumn{1}{c|}{\$ 100 / ton CH$_4$} & \multicolumn{1}{l|}{1,000 ton CH$_4$ / day} & \multicolumn{1}{c|}{1 day} \\ \cline{3-7} 
\multicolumn{1}{|c|}{} & \multicolumn{1}{l|}{} & \multicolumn{1}{l|}{Import Water} & \multicolumn{1}{c|}{-} & \multicolumn{1}{c|}{-} & \multicolumn{1}{l|}{30,000 ton H$_2$O / day} & \multicolumn{1}{c|}{0 days} \\ \cline{3-7} 
\multicolumn{1}{|c|}{} & \multicolumn{1}{l|}{} & \multicolumn{1}{l|}{Export Water} & \multicolumn{1}{c|}{-} & \multicolumn{1}{c|}{-} & \multicolumn{1}{l|}{30,000 ton H$_2$O / day} & \multicolumn{1}{c|}{0 days} \\ \cline{3-7} 
\multicolumn{1}{|c|}{} & \multicolumn{1}{l|}{} & \multicolumn{1}{l|}{Import Oxygen} & \multicolumn{1}{c|}{-} & \multicolumn{1}{c|}{-} & \multicolumn{1}{l|}{30,000 ton O$_2$ / day} & \multicolumn{1}{c|}{0 days} \\ \cline{3-7} 
\multicolumn{1}{|c|}{} & \multicolumn{1}{l|}{} & \multicolumn{1}{l|}{Export CO2} & \multicolumn{1}{c|}{-} & \multicolumn{1}{c|}{See Scenarios} & \multicolumn{1}{l|}{30,000 ton CO$_2$ / day} & \multicolumn{1}{c|}{0 days} \\ \cline{3-7} 
\multicolumn{1}{|c|}{} & \multicolumn{1}{l|}{} & \multicolumn{1}{l|}{Store Hydrogen} & \multicolumn{1}{c|}{-} & \multicolumn{1}{c|}{\$ 0.1 / ton H$_2$} & \multicolumn{1}{l|}{21,000 ton H$_2$ / day} & \multicolumn{1}{c|}{1 day} \\ \cline{3-7} 
\multicolumn{1}{|c|}{} & \multicolumn{1}{l|}{} & \multicolumn{1}{l|}{Store Natural Gas} & \multicolumn{1}{c|}{-} & \multicolumn{1}{c|}{\$ 0.1 / ton CH$_4$} & \multicolumn{1}{l|}{100,000 ton CH$_4$ / day} & \multicolumn{1}{c|}{1 day} \\ \hline
\multicolumn{7}{c}{} \\ \hline
\multicolumn{1}{|c|}{\multirow{2}{*}{3}} & \multicolumn{1}{l|}{\multirow{2}{*}{\begin{tabular}[c]{@{}l@{}}Ammonia\\ Production Facility\end{tabular}}} & \multicolumn{1}{l|}{Manufacture Ammonia} & \multicolumn{1}{c|}{-} & \multicolumn{1}{c|}{\$ 100 / ton H$_2$} & \multicolumn{1}{l|}{2,000 ton H$_2$ / day} & \multicolumn{1}{c|}{0 days} \\ \cline{3-7} 
\multicolumn{1}{|c|}{} & \multicolumn{1}{l|}{} & \multicolumn{1}{l|}{Store Hydrogen} & \multicolumn{1}{c|}{-} & \multicolumn{1}{c|}{\$ 0.1 / ton H$_2$} & \multicolumn{1}{l|}{21,000 ton H$_2$ / day} & \multicolumn{1}{c|}{1 day} \\ \hline
\multicolumn{7}{c}{} \\ \hline
\multicolumn{1}{|c|}{\multirow{9}{*}{4}} & \multicolumn{1}{l|}{\multirow{9}{*}{\begin{tabular}[c]{@{}l@{}}Hydrogen- and\\ Natural Gas-fired\\ Power Plant\end{tabular}}} & \multicolumn{1}{l|}{\begin{tabular}[c]{@{}l@{}}Burn Hydrogen to\\ Generate Electric Power\end{tabular}} & \multicolumn{1}{c|}{0.01 $\$^2$/ton H$_2$} & \multicolumn{1}{c|}{\$ 1000 / ton H$_2$} & \multicolumn{1}{l|}{1,000 ton H$_2$ / day} & \multicolumn{1}{c|}{1 day} \\ \cline{3-7} 
\multicolumn{1}{|c|}{} & \multicolumn{1}{l|}{} & \multicolumn{1}{l|}{\begin{tabular}[c]{@{}l@{}}Burn Natural Gas to\\ Generate Electric Power\end{tabular}} & \multicolumn{1}{c|}{0.01 $\$^2$/ton CH$_4$} & \multicolumn{1}{c|}{\$ 145 / ton CH$_4$} & \multicolumn{1}{l|}{3,000 ton CH$_4$ / day} & \multicolumn{1}{c|}{1 day} \\ \cline{3-7} 
\multicolumn{1}{|c|}{} & \multicolumn{1}{l|}{} & \multicolumn{1}{l|}{Consume Electric Power} & \multicolumn{1}{c|}{-} & \multicolumn{1}{c|}{-} & \multicolumn{1}{l|}{10,000 MWh / day} & \multicolumn{1}{c|}{0 days} \\ \cline{3-7} 
\multicolumn{1}{|c|}{} & \multicolumn{1}{l|}{} & \multicolumn{1}{l|}{Import Oxygen} & \multicolumn{1}{c|}{-} & \multicolumn{1}{c|}{-} & \multicolumn{1}{l|}{30,000 ton O$_2$ / day} & \multicolumn{1}{c|}{0 days} \\ \cline{3-7} 
\multicolumn{1}{|c|}{} & \multicolumn{1}{l|}{} & \multicolumn{1}{l|}{Export Water} & \multicolumn{1}{c|}{-} & \multicolumn{1}{c|}{-} & \multicolumn{1}{l|}{30,000 ton H$_2$O / day} & \multicolumn{1}{c|}{0 days} \\ \cline{3-7} 
\multicolumn{1}{|c|}{} & \multicolumn{1}{l|}{} & \multicolumn{1}{l|}{Export Heat Loss} & \multicolumn{1}{c|}{-} & \multicolumn{1}{c|}{-} & \multicolumn{1}{l|}{30,000 MMBTU / day} & \multicolumn{1}{c|}{0 days} \\ \cline{3-7} 
\multicolumn{1}{|c|}{} & \multicolumn{1}{l|}{} & \multicolumn{1}{l|}{Export CO2} & \multicolumn{1}{c|}{-} & \multicolumn{1}{c|}{See Scenarios} & \multicolumn{1}{l|}{30,000 ton CO$_2$ / day} & \multicolumn{1}{c|}{0 days} \\ \cline{3-7} 
\multicolumn{1}{|c|}{} & \multicolumn{1}{l|}{} & \multicolumn{1}{l|}{Store Hydrogen} & \multicolumn{1}{c|}{-} & \multicolumn{1}{c|}{\$ 0.1 / ton H$_2$} & \multicolumn{1}{l|}{21,000 ton H$_2$ / day} & \multicolumn{1}{c|}{1 day} \\ \cline{3-7} 
\multicolumn{1}{|c|}{} & \multicolumn{1}{l|}{} & \multicolumn{1}{l|}{Store Natural Gas} & \multicolumn{1}{c|}{-} & \multicolumn{1}{c|}{\$ 0.1 / ton CH$_4$} & \multicolumn{1}{l|}{100,000 ton CH$_4$ / day} & \multicolumn{1}{c|}{1 day} \\ \hline
\multicolumn{7}{c}{} \\ \hline
\multicolumn{1}{|c|}{\multirow{8}{*}{5}} & \multicolumn{1}{l|}{\multirow{8}{*}{\begin{tabular}[c]{@{}l@{}}Hydrogen- and\\ Natural Gas-fired\\ Steel Mill\end{tabular}}} & \multicolumn{1}{l|}{\begin{tabular}[c]{@{}l@{}}Burn Hydrogen to\\ Generate Industrial Heat\end{tabular}} & \multicolumn{1}{c|}{-} & \multicolumn{1}{c|}{\$ 300 / ton H$_2$} & \multicolumn{1}{l|}{1,000 ton H$_2$ / day} & \multicolumn{1}{c|}{1 day} \\ \cline{3-7} 
\multicolumn{1}{|c|}{} & \multicolumn{1}{l|}{} & \multicolumn{1}{l|}{\begin{tabular}[c]{@{}l@{}}Burn Natural Gas to\\ Generate Industrial Heat\end{tabular}} & \multicolumn{1}{c|}{-} & \multicolumn{1}{c|}{\$ 100 / ton CH$_4$} & \multicolumn{1}{l|}{1,000 ton CH$_4$ / day} & \multicolumn{1}{c|}{1 day} \\ \cline{3-7} 
\multicolumn{1}{|c|}{} & \multicolumn{1}{l|}{} & \multicolumn{1}{l|}{Consume Industrial Heat} & \multicolumn{1}{c|}{-} & \multicolumn{1}{c|}{-} & \multicolumn{1}{l|}{5,000 MMBTU / day} & \multicolumn{1}{c|}{0 days} \\ \cline{3-7} 
\multicolumn{1}{|c|}{} & \multicolumn{1}{l|}{} & \multicolumn{1}{l|}{Export Water} & \multicolumn{1}{c|}{-} & \multicolumn{1}{c|}{-} & \multicolumn{1}{l|}{30,000 ton H$_2$O / day} & \multicolumn{1}{c|}{0 days} \\ \cline{3-7} 
\multicolumn{1}{|c|}{} & \multicolumn{1}{l|}{} & \multicolumn{1}{l|}{Export CO2} & \multicolumn{1}{c|}{-} & \multicolumn{1}{c|}{See Scenarios} & \multicolumn{1}{l|}{30,000 ton CO$_2$ / day} & \multicolumn{1}{c|}{0 days} \\ \cline{3-7} 
\multicolumn{1}{|c|}{} & \multicolumn{1}{l|}{} & \multicolumn{1}{l|}{Import Oxygen} & \multicolumn{1}{c|}{-} & \multicolumn{1}{c|}{-} & \multicolumn{1}{l|}{30,000 ton O$_2$ / day} & \multicolumn{1}{c|}{0 days} \\ \cline{3-7} 
\multicolumn{1}{|c|}{} & \multicolumn{1}{l|}{} & \multicolumn{1}{l|}{Store Hydrogen} & \multicolumn{1}{c|}{-} & \multicolumn{1}{c|}{\$ 0.1 / ton H$_2$} & \multicolumn{1}{l|}{21,000 ton H$_2$ / day} & \multicolumn{1}{c|}{1 day} \\ \cline{3-7} 
\multicolumn{1}{|c|}{} & \multicolumn{1}{l|}{} & \multicolumn{1}{l|}{Store Natural Gas} & \multicolumn{1}{c|}{-} & \multicolumn{1}{c|}{\$ 0.1 / ton CH$_4$} & \multicolumn{1}{l|}{100,000 ton CH$_4$ / day} & \multicolumn{1}{c|}{1 day} \\ \hline
\multicolumn{7}{c}{} \\ \hline
\multicolumn{1}{|c|}{6} & \multicolumn{1}{l|}{\begin{tabular}[c]{@{}l@{}}Natural Gas\\ Import Station\end{tabular}} & \multicolumn{1}{l|}{Import Natural Gas} & \multicolumn{1}{c|}{-} & \multicolumn{1}{c|}{\$ 130 / ton CH$_4$} & \multicolumn{1}{l|}{100,000 ton CH$_4$ / day} & \multicolumn{1}{c|}{0 days} \\ \hline
\multicolumn{7}{c}{} \\ \hline
\multicolumn{1}{|c|}{\multirow{2}{*}{7}} & \multicolumn{1}{l|}{\multirow{2}{*}{LNG Terminal}} & \multicolumn{1}{l|}{Regasify Natural Gas} & \multicolumn{1}{c|}{-} & \multicolumn{1}{c|}{\$ 210 / ton CH$_4$} & \multicolumn{1}{l|}{100,000 ton CH$_4$ / day} & \multicolumn{1}{c|}{0 days} \\ \cline{3-7} 
\multicolumn{1}{|c|}{} & \multicolumn{1}{l|}{} & \multicolumn{1}{l|}{Store Natural Gas} & \multicolumn{1}{c|}{-} & \multicolumn{1}{c|}{\$ 0.1 / ton CH$_4$} & \multicolumn{1}{l|}{100,000 ton CH$_4$ / day} & \multicolumn{1}{c|}{1 day} \\ \hline
\multicolumn{7}{c}{} \\ \hline
\multicolumn{1}{|c|}{8} & \multicolumn{1}{l|}{\begin{tabular}[c]{@{}l@{}}Hydrogen Import\\ Station\end{tabular}} & \multicolumn{1}{l|}{Import Hydrogen} & \multicolumn{1}{c|}{-} & \multicolumn{1}{c|}{\$ 3000 / ton H$_2$} & \multicolumn{1}{l|}{100,000 ton H$_2$ / day} & \multicolumn{1}{c|}{0 days} \\ \hline
\multicolumn{7}{c}{} \\ \hline
\multicolumn{1}{|c|}{9} & \multicolumn{1}{l|}{\begin{tabular}[c]{@{}l@{}}Hydrogen Pipe\\ Line Intersection\end{tabular}} & \multicolumn{1}{l|}{Store Hydrogen} & \multicolumn{1}{c|}{-} & \multicolumn{1}{c|}{\$ 0.1 / ton H$_2$} & \multicolumn{1}{l|}{21,000 ton H$_2$ / day} & \multicolumn{1}{c|}{1 day} \\ \hline
\multicolumn{7}{c}{} \\ \hline
\multicolumn{1}{|c|}{10} & \multicolumn{1}{l|}{\begin{tabular}[c]{@{}l@{}}Natural Gas Pipe\\ Line Intersection\end{tabular}} & \multicolumn{1}{l|}{Store Natural Gas} & \multicolumn{1}{c|}{-} & \multicolumn{1}{c|}{\$ 0.1 / ton CH$_4$} & \multicolumn{1}{l|}{100,000 ton CH$_4$ / day} & \multicolumn{1}{c|}{1 day} \\ \hline
\multicolumn{7}{c}{} \\ \hline
\multicolumn{1}{|c|}{} & \multicolumn{1}{l|}{\begin{tabular}[c]{@{}l@{}}Hydrogen Pipe\\ Line\end{tabular}} & \multicolumn{1}{l|}{Transport Hydrogen} & \multicolumn{1}{c|}{-} & \multicolumn{1}{c|}{\$ 0.01 / ton H$_2$} & \multicolumn{1}{l|}{10,000 ton H$_2$ / day} & \multicolumn{1}{c|}{1 day} \\ \hline
\multicolumn{1}{|c|}{} & \multicolumn{1}{l|}{\begin{tabular}[c]{@{}l@{}}Hydrogen Pipe\\ Line 4\end{tabular}} & \multicolumn{1}{l|}{Transport Hydrogen} & \multicolumn{1}{c|}{-} & \multicolumn{1}{c|}{\$ 0.01 / ton H$_2$} & \multicolumn{1}{l|}{260 ton H$_2$ / day} & \multicolumn{1}{c|}{1 day} \\ \hline
\multicolumn{1}{|c|}{} & \multicolumn{1}{l|}{\begin{tabular}[c]{@{}l@{}}Hydrogen Pipe\\ Line 6\end{tabular}} & \multicolumn{1}{l|}{Transport Hydrogen} & \multicolumn{1}{c|}{-} & \multicolumn{1}{c|}{\$ 0.01 / ton H$_2$} & \multicolumn{1}{l|}{260 ton H$_2$ / day} & \multicolumn{1}{c|}{1 day} \\ \hline
\multicolumn{7}{c}{} \\ \hline
\multicolumn{1}{|c|}{} & \multicolumn{1}{l|}{\begin{tabular}[c]{@{}l@{}}Natural Gas\\ Pipe Line\end{tabular}} & \multicolumn{1}{l|}{Transport Natural Gas} & \multicolumn{1}{c|}{-} & \multicolumn{1}{c|}{\$ 0.01 / ton CH$_4$} & \multicolumn{1}{l|}{10,000 ton CH$_4$ / day} & \multicolumn{1}{c|}{1 day} \\ \hline
\end{tabular}
\end{table*}

\begin{table}[]
\centering
\tiny
\caption{Overview of the test case supply and demand data.}
\label{CH6:tab:testcaseSupplyDemand}
\begin{tabular}{cccccc}
\hline
\multirow{2}{*}{Day} & \multicolumn{2}{c}{\begin{tabular}[c]{@{}c@{}}Electric Power\\ Supply at Node 1 \\ {[}MWh / day{]}\end{tabular}} & \multirow{2}{*}{\begin{tabular}[c]{@{}c@{}}Hydrogen\\ Consumption for\\ Ammonia Production\\ at Node 3 {[}ton / day{]}\end{tabular}} & \multirow{2}{*}{\begin{tabular}[c]{@{}c@{}}Electric Power \\ Consumption\\ at Node 4 \\ {[}MWh / day{]}\end{tabular}} & \multirow{2}{*}{\begin{tabular}[c]{@{}c@{}}Industrial Heat \\ Consumption\\ at Node 5 \\ {[}MMBTU / day{]}\end{tabular}} \\ \cline{2-3}
 & \begin{tabular}[c]{@{}c@{}}Scenarios \\ 1 \& 3\end{tabular} & \begin{tabular}[c]{@{}c@{}}Scenarios \\ 2 \& 4\end{tabular} &  &  &  \\ \hline
1 & 0 & 6000 & 0 & 0 & 0 \\ \hline
2 & 0 & 6000 & 0 & 0 & 0 \\ \hline
3 & 0 & 6000 & 0 & 0 & 0 \\ \hline
4 & 0 & 6000 & 0 & 0 & 0 \\ \hline
5 & 0 & 6000 & 126 & 1435 & 35000 \\ \hline
6 & 0 & 6000 & 126 & 1459 & 35000 \\ \hline
7 & 0 & 6000 & 126 & 1312 & 35000 \\ \hline
8 & 0 & 6000 & 126 & 1189 & 35000 \\ \hline
9 & 0 & 6000 & 126 & 1402 & 35000 \\ \hline
10 & 0 & 6000 & 126 & 1404 & 35000 \\ \hline
11 & 0 & 6000 & 126 & 1363 & 35000 \\ \hline
12 & 0 & 6000 & 126 & 1416 & 35000 \\ \hline
13 & 0 & 6000 & 126 & 1479 & 35000 \\ \hline
14 & 0 & 6000 & 126 & 1288 & 35000 \\ \hline
15 & 0 & 6000 & 126 & 1281 & 35000 \\ \hline
16 & 0 & 0 & 126 & 1455 & 35000 \\ \hline
17 & 0 & 0 & 126 & 1480 & 35000 \\ \hline
18 & 0 & 0 & 126 & 1476 & 35000 \\ \hline
19 & 0 & 0 & 126 & 1275 & 35000 \\ \hline
20 & 0 & 0 & 0 & 0 & 0 \\ \hline
\end{tabular}
\end{table}

This subsection first introduces the physical lay-out of the test case. Then, it discusses the device models for the processes. 

\subsubsection{Test Case Physical Lay-out}
The lay-out of the test case is derived from the topology of the Dutch industrial and infrastructure clusters (see Fig. \ref{CH6:fig:testcaseNetwork}): 

The \emph{south-west area} of The Netherlands accommodates critical energy infrastructure: a hydrogen electrolysis facility (Node 1), a steam-methane reformation facility (Node 2), a power generation cluster (Node 4), and an LNG terminal (Node 7). The \emph{north-west region} of the test case contains heavy industry: a steel mill that uses a combination of natural gas and hydrogen as its fuel (Node 5). The \emph{north-east} contains infrastructure that imports natural gas (Node 6) and hydrogen (Node 8) to the system. Finally, the \emph{mid- and south-east} region contains two pipeline junctions (Nodes 9 and 10) and an ammonia factory (Node 3).

The system consists of industrial clusters, connected through dedicated pipelines for hydrogen and natural gas. Table \ref{CH6:CH6:tab:testcaseResources} provides an overview of the clusters with the associated processes, the cost, the capacities of the processes, and the processing time. Note that most process capacities are not intended to be a binding constraint, however, the capacities of hydrogen pipe lines 4 and 6 are likely to be binding in some scenarios.

\subsubsection{Device Models}
The dynamics of the test case processes are described through their device models. These device models are (mass-based) ratios between input and output operands derived from their stoichiometry. All weights are in metric ton (1000 kg). The device models of the transformative processes are derived from the relevant literature:
\begin{enumerate}
    \item Electrolyze Water to Hydrogen and Oxygen \cite{Scott:2019:00}: 
    \begin{equation}
        2 H_2O + \text{ Electric Power } \rightarrow 2 H_2 + O_2
    \end{equation}
    This process consumes 40 - 50 MWh / ton $H_2$ \cite{Bertuccioli:2014:00}. The associated mass-based ratio is:
    \begin{align}\nonumber 
        8.936\ & ton\ H_2O\ +\ 40\ MWh\ \rightarrow \\ & 1\ ton\ H_2\ +\ 7.936\ ton\ O_2
    \end{align}
    \item Reform Steam and Methane to Hydrogen and CO$_2$:
    The stoichiometric equation combines the steam reformation process and the shift reaction\cite{Rosen:1991:00,Peng:2012:01}:
    \begin{equation}
        CH_4 + 2 H_2O + \text{Industrial Heat} \rightarrow 4 H_2 + CO_2
    \end{equation}
    This process consumes around 19.4 MMBTU per ton H$_2$ produced. The associated mass-based stochiometric equation for steam methane reformation is:
    \begin{align}\nonumber 
        1.989\ & ton\ CH_4 +\ 4.468\ ton\ H_2O +\ 19.4\ \text{MMBTU} \\ &\rightarrow 1\ ton\ H_2\ +\ 5.457\ ton\ CO_2
    \end{align}
    \item Burn Natural Gas to Generate Industrial Heat:
    \begin{equation}
        CH_4 + 2O_2 \rightarrow 2 H_2O + CO_2 + \text{ Industrial Heat}
    \end{equation}
    For this ratio, it is assumed that all generated industrial heat is used productively (with a HHV of $CH_4$ of 891 kJ / mol)  \cite{National-Academy-of-Engineering:2004:00}. The associated mass-based ratio is:
    \begin{align}\nonumber 
        1\ & ton\ CH_4\ +\ 3.989\ ton\ O_2\ \rightarrow \\ \nonumber 
        & 2.246\ ton\ H_2O\ +\ 2.743\ ton\ CO_2\ +\  \\ 
        & 52.6 \text{ MMBTU} 
    \end{align} 
    \item Burn Natural Gas to Generate Electric Power:
    \begin{align}\nonumber 
        CH_4 + 2O_2 & \rightarrow 2H_2O + CO_2 + \\
        & \text{ Electric Power} + \text{ Heat Loss}
    \end{align}
    The heat rate is assumed at 7633 BTU / kWh \cite{National-Academy-of-Engineering:2004:00,Energy-Information-Administration:2020:00}. The associated mass-based ratio is:
    \begin{align}\nonumber 
        1\ & ton\ CH_4\ +\ 3.989\ ton\ O_2\ \rightarrow \\ 
        & 2.246\ ton\ H_2O\ +\ 2.743\ ton\ CO_2\ +\ \nonumber \\ 
        & 6.897 \text{ MWh}\ +\ 29.1 \text{ MMBTU}
    \end{align} 
    \item Burn Hydrogen to Generate Industrial Heat:
    \begin{equation}
        2H_2 + O_2 \rightarrow 2H_2O + \text{ Industrial Heat}
    \end{equation}
    Where all generated heat is used productively \cite{National-Academy-of-Engineering:2004:00}. The associated mass-based ratio is:
    \begin{align}\nonumber 
        1\ & ton\ H_2\ +\ 7.936\ ton\ O_2\ \rightarrow \\ & 8.936\ ton\ H_2O\ +\ 134.5\ \text{MMBTU}
    \end{align} 
    \item Burn Hydrogen to Generate Electric Power:
    \begin{equation}
        2H_2 + O_2 \rightarrow 2H_2O + \text{ Electric Power} + \text{ Heat Loss}
    \end{equation}
    For this ratio, the heat rate of the hydrogen-fired turbine is assumed to be 7633 BTU / kWh. As a result, the mass-based ratio is:
    \begin{align}\nonumber 
        1\ & ton\ H_2\ +\ 7.936\ ton\ O_2\ \rightarrow \\ & 8.936\ ton\ H_2O\ + 17.616 \text{ MWh}\ +\ \nonumber \\ 
        &  74.3 \text{ MMBTU}
    \end{align} 
\end{enumerate}
\begin{figure}[t]
\includegraphics[width=3.2in]{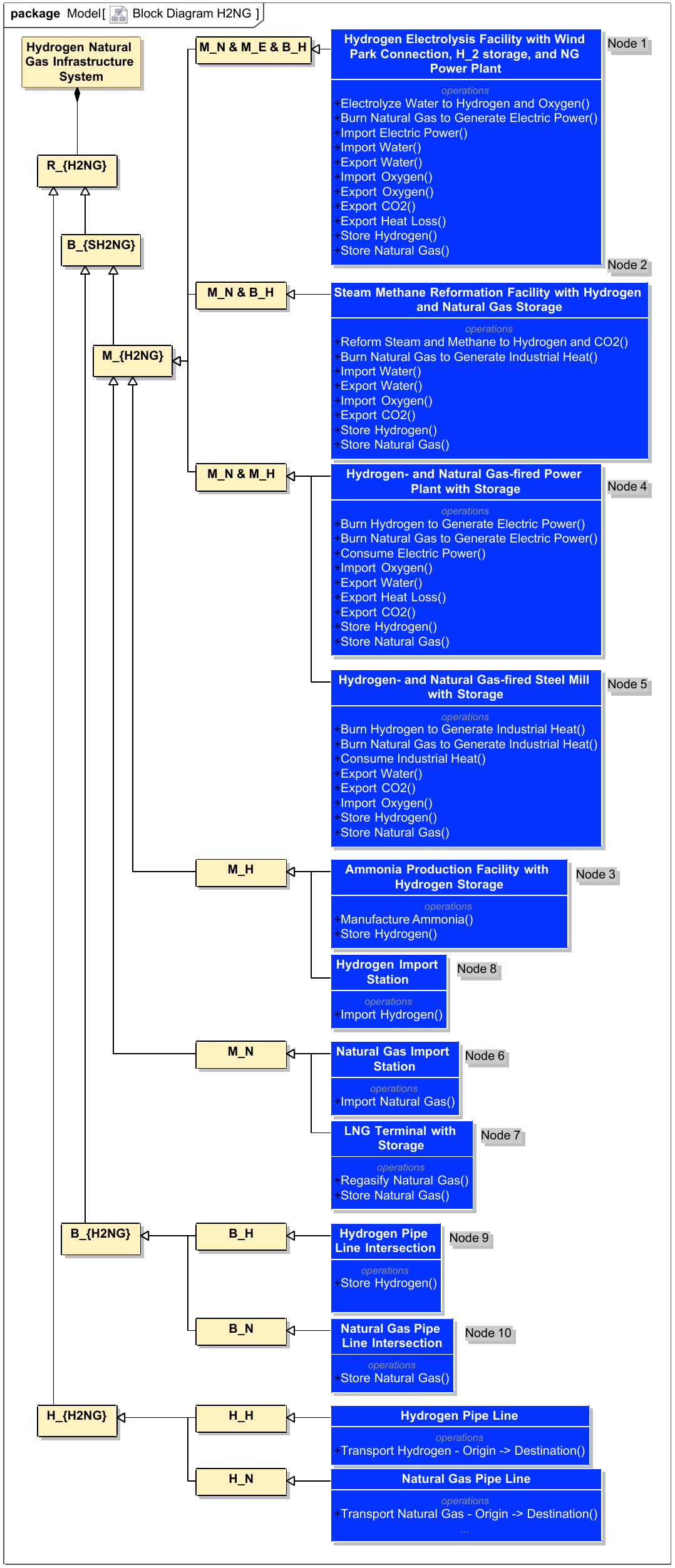}
\caption{SysML Block Diagram of the Hydrogen-Natural Gas System Resources.}
\label{CH6:fig:bd:H2NG}
\end{figure}

\begin{figure*}[t!]
\centering
\includegraphics[width=6in]{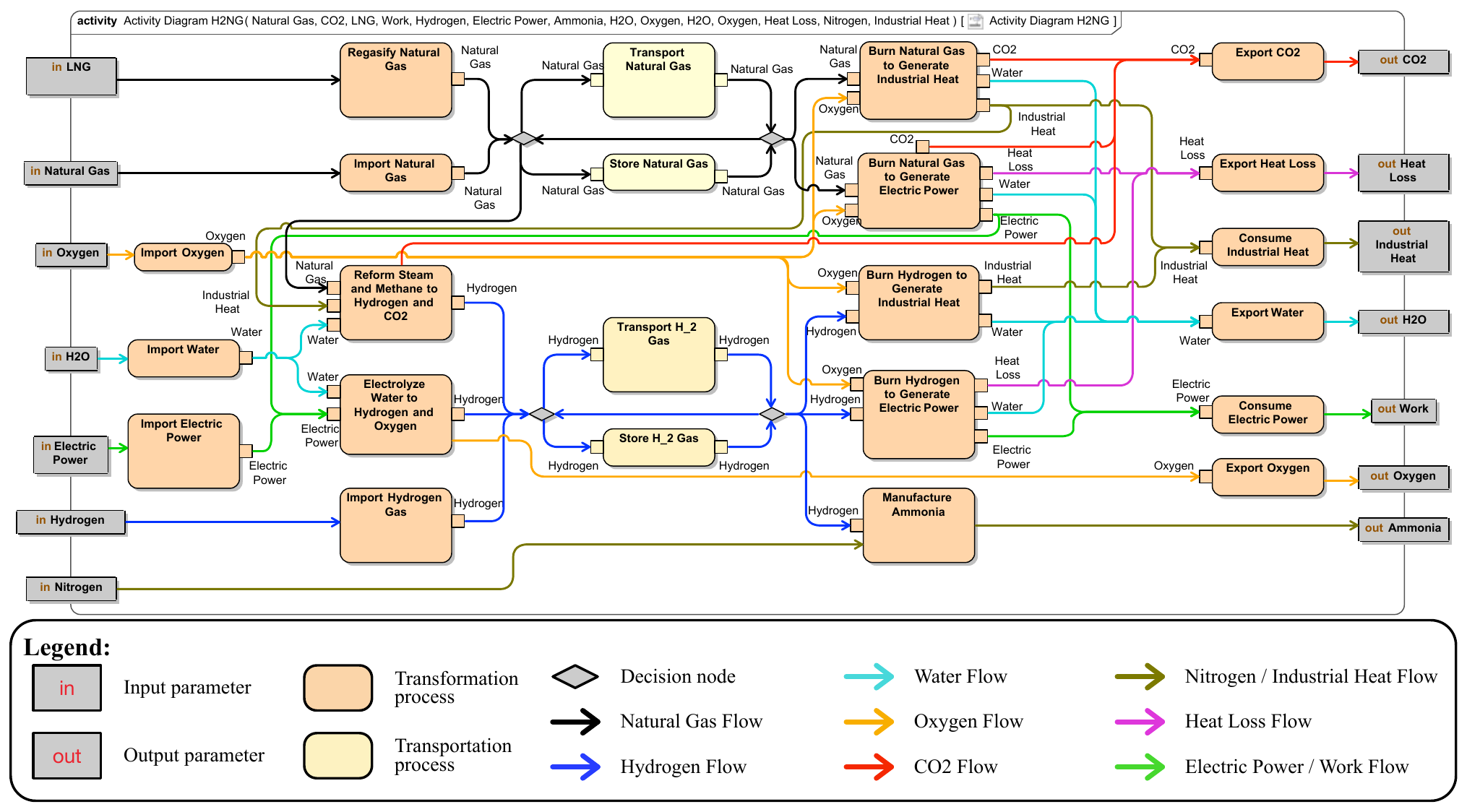}
\caption{SysML Activity Diagram of the Hydrogen-Natural Gas System Processes.}
\label{CH6:fig:ad:H2NG}
\end{figure*}

The remaining transformation processes import or consume operands and are defined only what they bring into or take out of the system, as displayed in Table \ref{CH6:CH6:tab:testcaseResources}. 

Finally, the test case assumes that all transportation processes are lossless:
\begin{itemize}
    \item Transport Natural Gas, expressed in ton per day. 
    \item Transport Hydrogen, expressed in ton per day.
\end{itemize}

Table \ref{CH6:tab:testcaseSupplyDemand} presents the four \emph{supply and demand} curves. Note that in this test case, electric power cannot be stored and needs to be used immediately.

\vspace{-0.1in}
\subsection{Scenario Data}\label{CH6:subsec:IE:scenarios}
The test case optimizes four scenarios:
\begin{itemize}
\item Scenario 1: the base case scenario without carbon pricing or a fixed renewable electricity supply.
\item Scenario 2: incorporates carbon pricing of \$250 per ton for carbon emissions at the steel mill. It does not include a fixed renewable electricity supply.
\item Scenario 3: introduces the fixed renewable electricity supply. It does not include carbon pricing.
\item Scenario 4: incorporates carbon pricing of \$500 per ton for all resources and the fixed renewable electricity supply.
\end{itemize}
For each of these scenarios, the goal is to have the lowest fulfillment cost for the three demand operands over the 20 day time horizon ($K = 20$). 
 
\vspace{-0.1in}
\section{Results and Discussion}\label{CH6:sec:Results}
This section applies the hetero-functional network minimum cost flow program to the hydrogen-natural gas test case. Sec. \ref{CH6:subsec:Results:HFGTstruc} first covers the hetero-functional graph theory structural model. Sec. \ref{CH6:subsec:Results:HFGTdynamics} then develops the dynamic model. Sec. \ref{CH6:subsec:Results:HFGTprogram} defines the optimization program. Finally, Sec. \ref{CH6:subsec:Results:HFGTresults} discusses the results of the optimization program. 
\vspace{-0.1in}
\subsection{Hetero-functional Graph Theory Structural Model}\label{CH6:subsec:Results:HFGTstruc}
The Hetero-functional Graph Theory structural model provides the foundation for the development of a dynamic model and an optimal control program.  It contains the System Concept (Sec. \ref{CH6:subsec:B:HFGTsysconcept}), the Hetero-functional Incidence Tensor (Sec. \ref{CH6:subsec:B:HFGTincidencetensor}), and the Service Model (Sec. \ref{CH6:subsec:B:HFGTservices}), which includes the Service Nets and the Service Feasibility Matrices.  To facilitate the reproducibility of the work, the sizes of these matrices are provided and the associated data sets are found in \cite{Schoonenberg:2021:ISC-TP02}.  

Fig. \ref{CH6:fig:bd:H2NG} describes the system resources with a SysML block definition diagram. The test case contains 27 resources of which 8 are transformation resources, 2 are independent buffers, and 17 are transportation resources. The diagram also shows the processes allocated to each of the resources.
\begin{figure}[t]
\centering
\includegraphics[width=3.2in]{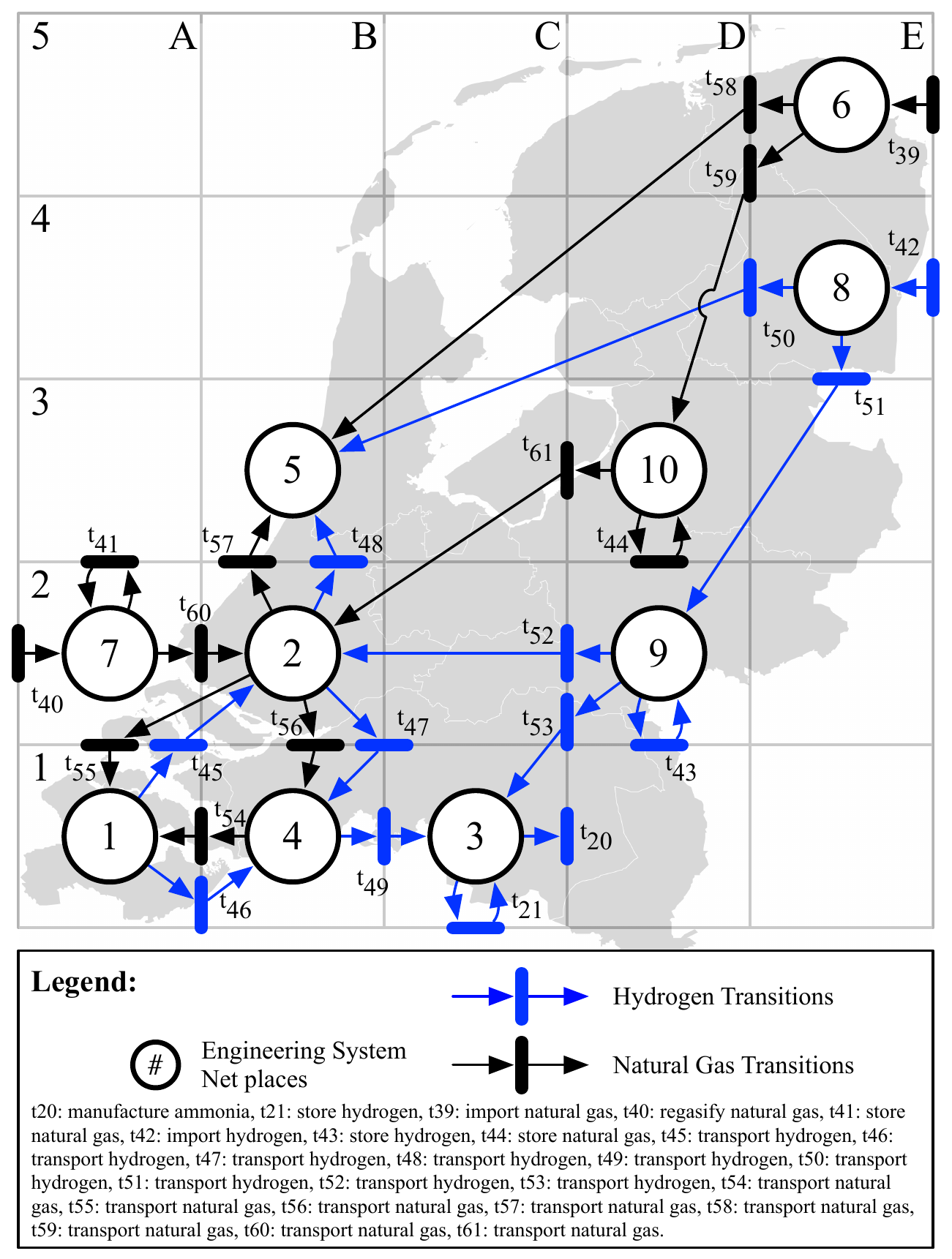}
\vspace{-0.1in}
\caption{The engineering system net. Places 1, 2, 4, and 5 are presented in more detail in Figure \ref{CH6:fig:CH6:ESNdetail}. The operand colors correspond to the the activity diagram (Fig. \ref{CH6:fig:ad:H2NG}).}
\label{CH6:fig:CH6:acCPN}
\end{figure}

\begin{figure}[t]
\centering
\includegraphics[width=3.2in]{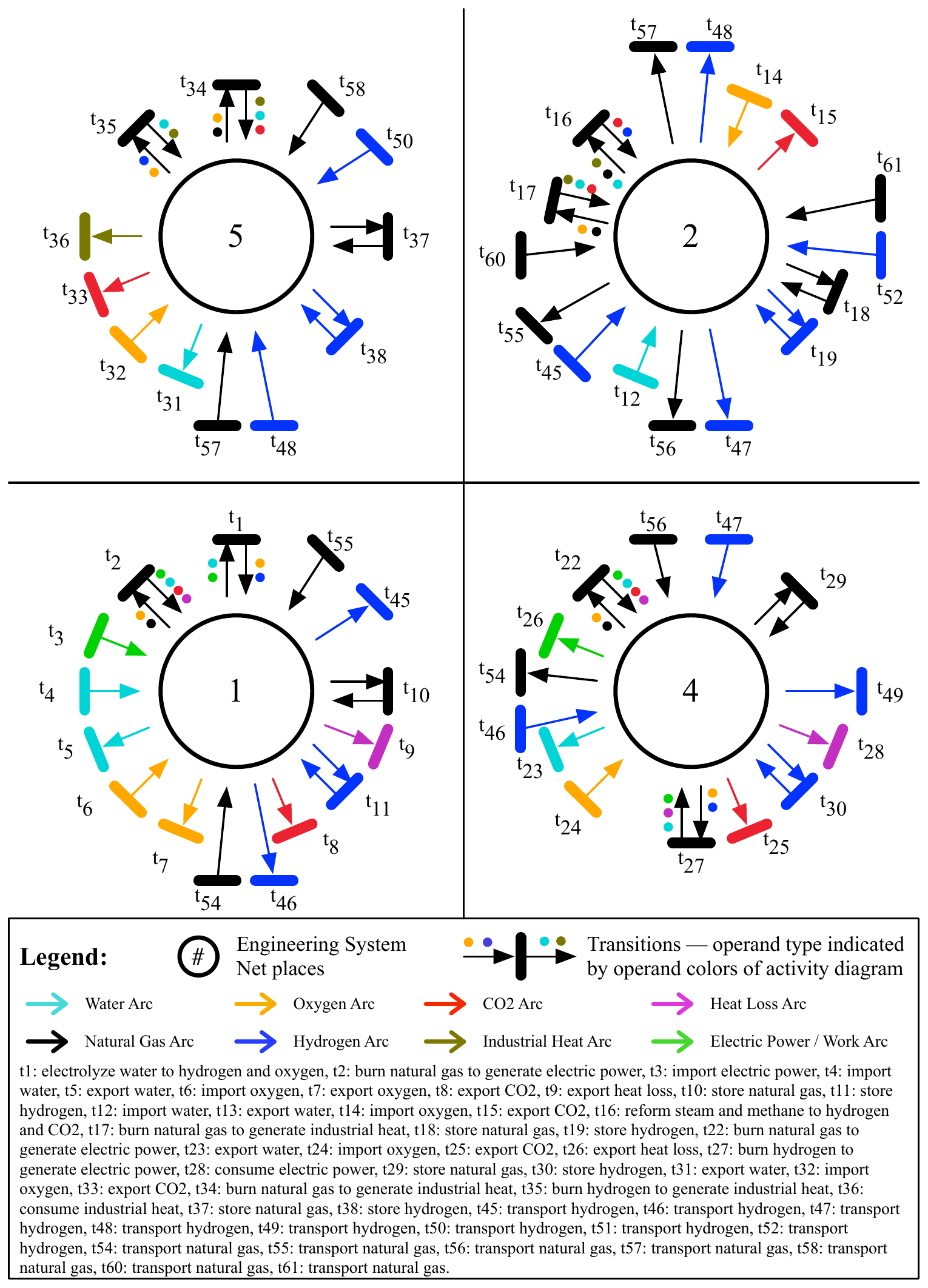}
\vspace{-0.1in}
\caption{A detailed picture of places 1, 2, 4, and 5 from the engineering system net as presented in Figure \ref{CH6:fig:CH6:acCPN}. The operand colors correspond to the the activity diagram (Fig. \ref{CH6:fig:ad:H2NG}). Transitions with multiple input/output operands use colored dots next to the associated arc to indicate the operand types.
}
\label{CH6:fig:CH6:ESNdetail}
\end{figure}

\begin{figure*}[t]
\centering
\includegraphics[width=5.5in]{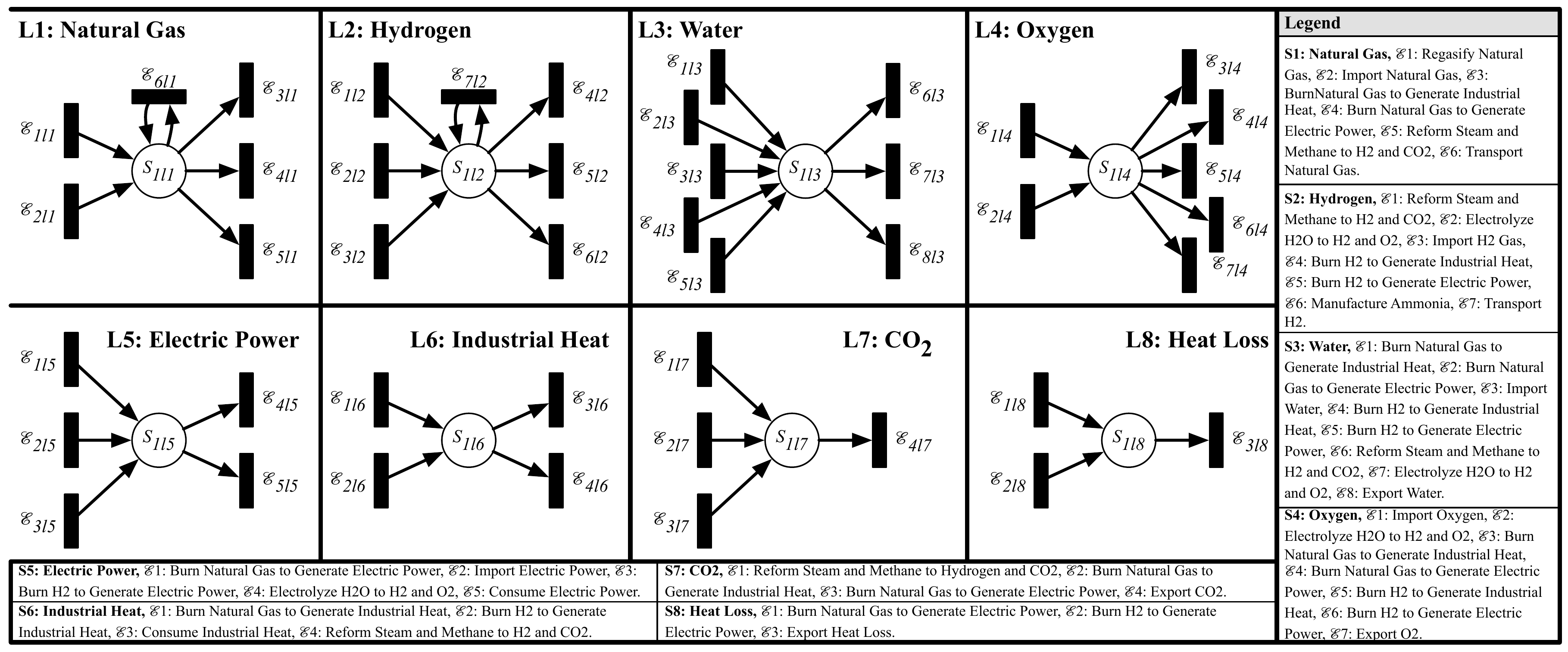}
\caption{Service nets for the Hydrogen Natural Gas test case. }
\label{CH6:fig:servicenets:H2NG}
\end{figure*}

Fig. \ref{CH6:fig:ad:H2NG} describes the system processes with a SysML activity diagram. The activity diagram shows the functional reference architecture and the feasible system process sequences.

The system concept, or the allocated architecture, maps the system processes onto the system resources with the knowledge base. As expected, the knowledge base has size $219 \times 27$, with 61 filled elements.

The hetero-functional incidence tensor describes the association of the system buffers with the capabilities and the system operands. It is defined in Definitions \ref{CH6:def:HFITneg} and \ref{CH6:def:HFITpos}. For this test case, the projected Hetero-functional Incidence Tensor has size: $8 \times 10 \times 61$ (operands by buffers by capabilities) and it has 98 filled elements. The associated Engineering System Net is presented in Figs. \ref{CH6:fig:CH6:acCPN} and \ref{CH6:fig:CH6:ESNdetail}, where the latter provides a detailed look at Nodes 1, 2, 4, and 5.

Fig. \ref{CH6:fig:servicenets:H2NG} describes the service nets for all eight operands in the system. 
The services are synchronized with the engineering system capabilities through the service feasibility matrix. 

\subsection{Hetero-functional Graph Theory Dynamic Model}\label{CH6:subsec:Results:HFGTdynamics}
The hetero-functional network dynamics model was introduced in Sec. \ref{CH6:sec:HFGTdynamics}. The first element of the dynamic model is the engineering system net, modified to incorporate the device models. The device model matrices $D_R^+$ and $D_R^-$ have size: $\sigma(L)\times \sigma(P) = 8 \times 219$. The incidence matrices in the engineering system net are modified as noted Eqs. \ref{CH6:CH6:eq:HFITwDMpos} and \ref{CH6:CH6:eq:HFITwDMneg}. The second element of the dynamic model contains the service nets. These are directly adopted from the structural model. The final element of the dynamic model describes the synchronization equations for the coupling of the engineering system net and the service nets. The synchronization matrices $\widehat{\Lambda}^+_i$ and $\widehat{\Lambda}^-_i$ are defined by incorporation of the device models in Eqs. \ref{CH6:CH6:eq:syncMatrixPos} and \ref{CH6:CH6:eq:syncMatrixNeg} and have the same size as the service feasibility matrices as defined in the previous section. 
\vspace{-0.15in}
\subsection{Hetero-functional Network Min. Cost Flow Program}\label{CH6:subsec:Results:HFGTprogram}
The definition of the quadratic program follows the description in Sec. \ref{CH6:sec:HFGTprogram} and more specifically Eq. \ref{ch6:eq:QPcanonicalform:1}-\ref{ch6:eq:QPcanonicalform:4}.

The decision vector $x$ has size: $8,463 \times 1$.  The quadratic cost-coefficient matrix $F_{QP}$ has size: $8,463 \times 8,463$.  The linear cost-coefficient matrix $f_{QP}$ has size: $8,463 \times 1$.  The linear equality constraint coefficient matrix $A_{QP}$ has size: $7,323 \times 8,463$.  The linear equality constraint vector  $B_{QP}$ has size: $7,323 \times 1$.  The linear inequality coefficient matrix $D_{QP}$ has size: $1,281 \times 8,463$.  Finally, the linear inequality constraint vector $E_{QP}$ has size: $1,281 \times 1$.

The quadratic cost function, the $F_{QP}$-matrix, has positive eigenvalues. The resulting mathematical program is a convex quadratic program. The linear equality constraints matrix $A_{QP}$ consists of block rows that reflect the equality constraints (as introduced in Sec. \ref{CH6:subsec:HFGTprog:full}). These block rows are now discussed in order.

\begin{table*}[t!]
\centering
\caption{Overview of cost and carbon emissions per scenario}
\label{CH6:tab:Results:overview}
\begin{tabular}{ccccc}
 & \textbf{Scenario 1} & \textbf{Scenario 2} & \textbf{Scenario 3} & \textbf{Scenario 4} \\ \hline
Total Cost & \$6,092,627.17 & \$10,452,421.24 & \$11,777,395.62 & \$25,244,985.80 \\ \hline 
Total CO$_2$ Emissions & 47,026.74 ton & 45,041.97 ton & 33,091.17 ton & 0 ton
\end{tabular}
\end{table*}

\vspace{-0.1in}
\subsection{Scenario Results}\label{CH6:subsec:Results:HFGTresults}

\begin{figure*}[t]
\centering
\includegraphics[width=6in]{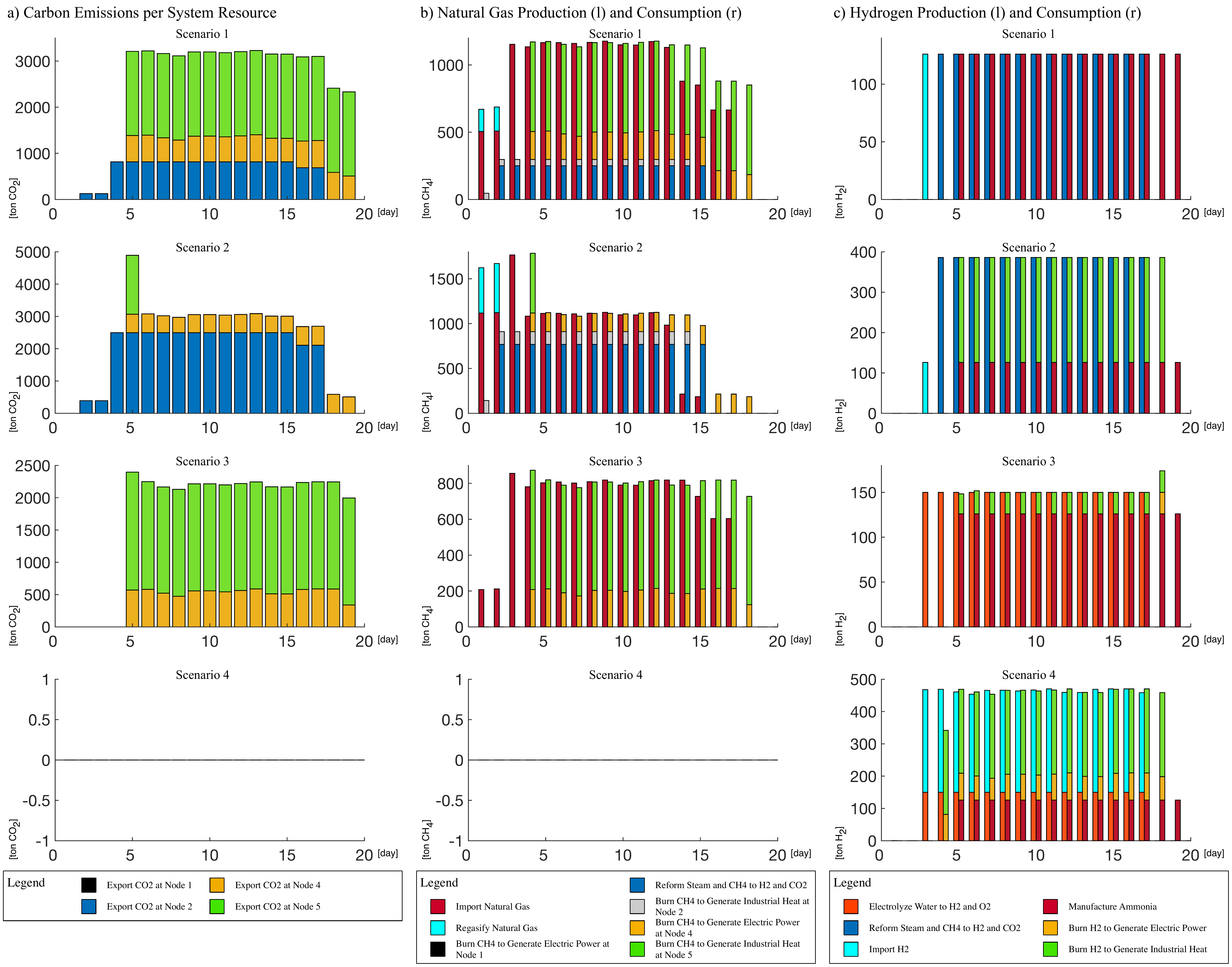}
\caption{a) Carbon emissions per system resource for all time steps. b) Natural Gas Production (left column) vs. Natural Gas Consumption (right column) for all time steps. c) Hydrogen Production (left column) vs. Hydrogen Consumption (right column) for all time steps. }
\label{CH6:fig:Results:Results}
\end{figure*}

The final results of this work encompass the optimization of the test case program for the four different scenarios. The optimization program matrices were defined in MATLAB 2019a and solved as a quadratic program using the CONOPT 3 solver in GAMS. All programs were found to be locally optimal in less than 2 seconds when running the program on a MacBook Pro (15-inch, 2017) with a 3.1 GHz Quad-Core Intel Core i7 and 16 GB RAM. 

Table \ref{CH6:tab:Results:overview} provides an overview of the total cost and the carbon emissions of each of the four scenarios. Fig. \ref{CH6:fig:Results:Results}a shows a breakdown of the carbon emissions per resource, Fig. \ref{CH6:fig:Results:Results}b the natural gas balance (the generation and consumption for each of the resources), and Fig. \ref{CH6:fig:Results:Results}c the hydrogen balance for the system as a whole. The results of the scenarios are now compared. 

Scenario 1 is the least expensive scenario, but it emits the highest level of carbon dioxide. Since there is no renewable energy input to the system, electrolysis is only used to replace steam reformation in time step 3. Steam reformation requires an extra time step to ramp up from a cold start and cannot fulfill the hydrogen demand in time step 4. The demand for industrial heat in the steel mill is satisfied by natural gas as the least cost option.

Scenario 2 imposes a carbon tax of \$ 250 per ton CO$_2$ emitted by the steel mill. Scenario 2 is 72\% more expensive than scenario 1, while emitting 4\% less carbon dioxide. The steel mill sources almost all of its industrial heat from hydrogen to avoid the carbon tax. Its hydrogen supply is produced by the SMR process and causes a substantial increase of carbon emissions at the SMR facility relative to scenario 1. Not all industrial heat is satisfied by hydrogen, as the capacity of Hydrogen Pipe Line 4 is insufficient. The remainder of the industrial heat is supplied by natural gas combustion, as the imported hydrogen is more expensive than the combination of imported natural gas and a carbon tax.

Scenario 3 incorporates a predetermined supply of renewable electricity to the system. The electricity cannot be transported and forces the production of hydrogen through electrolysis. The total carbon dioxide emissions are 30\% lower than in scenario 1. The total cost of scenario 3 is 93\% higher than scenario 1 and 13\% higher than scenario 2. The hydrogen through electrolysis is predominantly used to supply the ammonia facility and the left-overs are used to produce industrial heat in the steel mill. Natural gas is used to provide the bulk of the industrial heat in the steel mill as the least cost option. 

Scenario 4 combines the renewable electricity supply with a carbon tax of \$500 per ton CO$_2$ at all locations. This results in a cost increase of 314\% over scenario 1 and zero carbon emissions (within the boundaries of this system). As the use of natural gas is clearly too expensive in this scenario, the supply of hydrogen is satisfied by the least cost routing of the hydrogen. The steel mill is a single transportation process removed from the hydrogen import facility and therefore, it receives predominantly imported hydrogen. Hydrogen Pipe Line 6 reaches its capacity limit as a result. 

From the optimization results of these four scenarios, it is clear that the hetero-functional network minimum cost flow program enables the optimization of a continuous flow multi-operand system over time with storage of operands, transformation of operands, and the explicit description of the state of operands. This holistic program enables the user to study trade-offs and synergies in the behavior of interdependent systems. 
\vspace{-0.1in}

\section{Conclusion}\label{CH6:sec:Conclusion}
This work set out to define a hetero-functional network minimum cost flow optimization program that enables the optimization of large flexible engineering systems across multiple types of operands. This program is the first of its kind, as it is the first hetero-functional graph theory-based optimization program. 

In the process of developing the first hetero-functional network minimum cost flow optimization program, the work has established the first formal connection between the Hetero-functional Incidence Tensor, arc-constant Colored Petri nets, and the Engineering System Net. Furthermore, it has defined the first integration of device models to the feasibility matrices that couple the engineering system net and the system services net. Additionally, the implementation of the hetero-functional network minimum cost flow optimization program accommodates the explicit definition of time and therefore storage. Moreover, the program accommodates both linear and quadratic optimization of such a dynamic, hetero-functional network model. Finally, the demonstration of the hetero-functional network minimum cost flow program in this paper has lead to the definition of the first hydrogen-natural gas infrastructure test case. 
\vspace{-0.15in}
\bibliographystyle{IEEEtran}
\bibliography{LIINESLibrary}
\end{document}